\documentclass[final,3p,times,authoryear]{elsarticle}

\usepackage{graphicx} % Required for inserting imagesy
\usepackage{xcolor}
\usepackage{amsmath} 
\usepackage{amssymb}
\pagecolor{white}
\usepackage{parskip} % evitar sangria
\usepackage{hyperref}
\usepackage{natbib}  % bibliografia 
\usepackage{multicol}
\usepackage{ulem}

\journal{Astroparticle Physics}

\usepackage[utf8]{inputenc}
\usepackage{textcomp}

\begin{document}
\begin{frontmatter}

\title{No evidence for Keplerian taper of far-out galactic rotation \\ 
in the SPARC galaxy database}
\author{Adriana Bariego-Quintana$^a$ and  Felipe  J. Llanes-Estrada$^b$}
\affiliation[IFIC]{organization={Instituto de Fisica Corpuscular},
            city={Paterna},
            postcode={E-46980}, 
            state={Valencia},
            country={Spain}}
            
\affiliation[UCM]{
organization={Dept. Fisica Teorica \& IPARCOS, Univ. Complutense de Madrid},%Department and Organization
            city={Madrid},
            postcode={E-28040}, 
            state={},
            country={Spain} }

%\justifying
% \maketitle

\begin{abstract}
    We present a statistical analysis of the 175 SPARC galactic rotation curves to test the hypothesis of whether the Keplerian velocity tapering  at large radii ($V(r)\propto 1/\sqrt{r}$) germane to a convergent mass distribution \textcolor{black}{in typical spherical halo models} agrees with observational data.  
    The null hypothesis is Rubin's flat-rotation curve, $V(r)=\text{constant}$ -such as can be obtained from a spherical, isothermal-like density profile, or alternatively with a very prolate halo-. 
    To decide whether we adopt the null (Rubin behaviour) or alternative (\textcolor{black}{Keplerian} behaviour) hypothesis, we evaluate the derivative in each galaxy of $V(r)$ with its last data points. The test is model independent inasmuch we are testing for the \emph{slope} of the \textcolor{black}{dark matter} rotation curve, whether it is or not compatible with zero.
    We conclude that the data is presently compatible with the null hypothesis -no taper off, no decline of $V(r)$ is seen.
    Separately, beyond SPARC, our own Milky Way galaxy, for which recent data sets have been reported, does show clear $V(r)$ fall-off at the level of 20\%.
\end{abstract}

\end{frontmatter}

%%%%%%%%%%%%%%%%%%%%%%%%%%%%%%%%%%%%%%%%%%%%%%%%%%%%%%%%%%%%%%%%%%%5
\section{Introduction}
%%%%%%%%%%%%%%%%%%%%%%%%%%%%%%%%%%%%%%%%%%%%%%%%%%%%%%%%%%%%%%%%%%%

\begin{figure}[h!]
    \centering
    \includegraphics[width=0.56\linewidth]{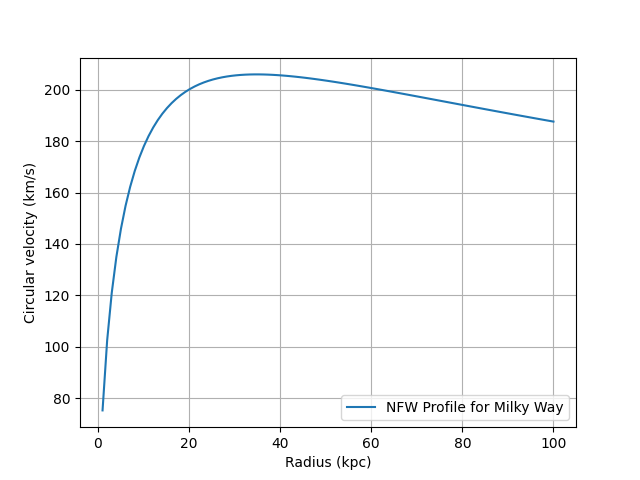}
    \includegraphics[width=0.4\linewidth]{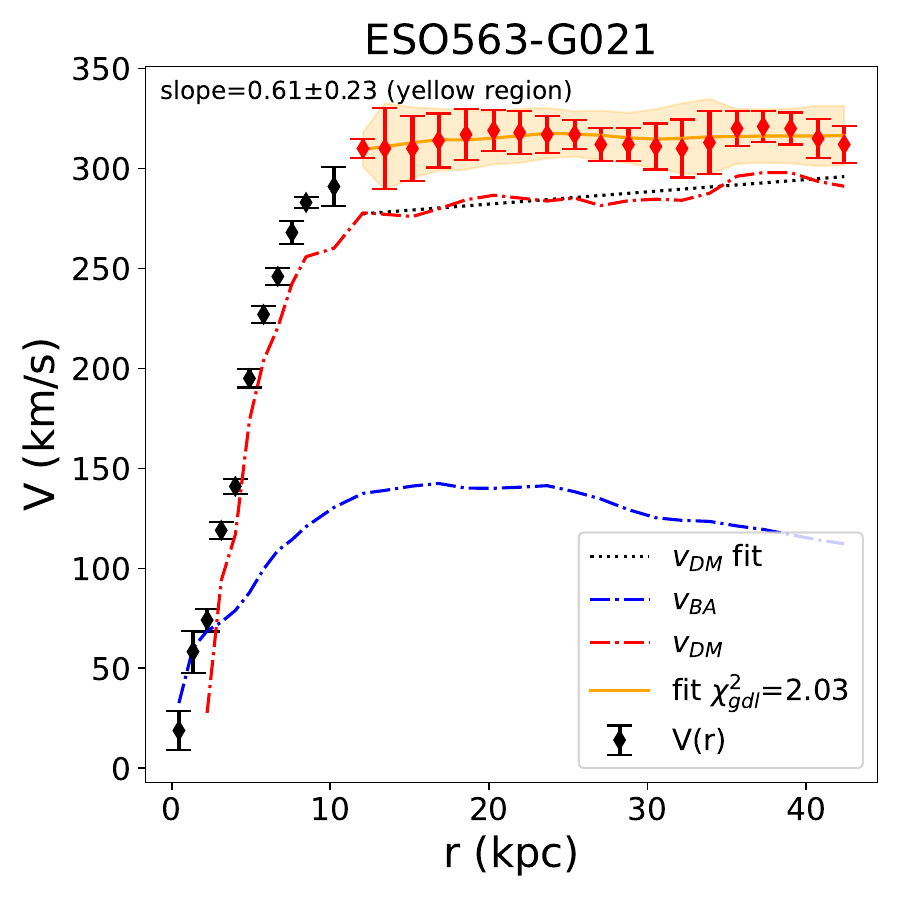}
    \caption{{\bf Left:} Dark-Matter induced rotation curve within a supposedly spherical NFW halo (with the parameters of the Milky Way \textcolor{black}{to fix the units, though the shape is universal})
    At distances from the center doubling that of the local maximum, the velocity curve clearly displays the Keplerian $1/\sqrt{r}$ fall-off. {\bf Right:} Example galaxy with well measured rotation curve that does not fall-off but rather flattens out (shaded area) at large radius (even past three to four times the would-be maximum) in contrast to the NFW prediction. This apparent shape disagreement motivates our study. }
    \label{fig:NFW}
\end{figure}

There is a trove of observational evidence pointing to an unidentified gravitating component in the Universe \citep{1998CeMDA..72...69B,profumo_intro_dark_matter}, all the way from galactic to cluster to cosmological scales: even the power spectrum of the Cosmic Background radiation signals the need for such an extra component~\citep{Planck_2020}, also necessary to explain the still ongoing large scale structure formation.

While extensions of gravitational theories \citep{1983} are still discussed in the literature, a leading hypothesis posits that large-scale observations are best explained by a new component of matter, Dark Matter (DM) \citep{Bertone_2018}.
According to Cosmic Structure formation, DM started seeding overdensities already in the radiation-dominated era and created large potential wells which over the matter-dominated expansion pulled baryonic matter in and started structure formation~\citep{1993sfu..book.....P,2006}.
Cosmological simulations suggest that in the present universe galaxies are surrounded by DM haloes that alter probe-mass trajectories. 

Without speculating about the nature of this gravitating component, one can aim to constrain the potential of DM haloes to understand their shape, distribution, and underlying properties~\citep{Bariego-Quintana:2024uaw}, for example from stellar streams~\citep{Walder:2024zfv}. Of the abundant observational evidence supporting the existence of DM, the indefinitely flat rotation curve seen in a large fraction of galaxies --as studied in depth by V. Rubin and collaborators~\citep{1978} or by Bosma~\citep{1978PhDT.......195B} in  the 1970s--  stands out for its simplicity.

Standard theoretical analysis assumes a spherical DM halo.  Fitting the matter concentrations in cosmological simulations~\citep{DiCintio:2013qxa} leads to functions such as 
the \textcolor{black}{profiles of Einasto, Moore, Burkert, ... or the very common, and that we take as illustrative example,} Navarro-Frenk-White (NFW)~\citep{1997ApJ...490..493N} one which can be written as the following density profile
\begin{equation}
     \rho_{NFW} =  \frac{\rho_0}{x \cdot(1+x)^2}
 \end{equation}
where $x=r/r_s$ is the distance at which d$\log{\rho}/$d$\log{r}=-2$, the slope which actually supports flat rotation curves\footnote{ For example, for the Milky Way, traditional analysis~\citep{Fabrizio_Nesti_2013} suggested $r_s=16.1^{+17}_{-7.8}$ kpc 
and $\rho_0=1.4^{+2.9}_{-0.93}\cdot 10^{-7}$ $M_\odot/$kpc$^3$ is the local DM density at about the solar system.}.
 The rotation curve $V(r)$ for such a spherical NFW halo is easily derived from orbital equilibrium,
\begin{eqnarray}
    V(r) &=& \sqrt{\frac{4\pi G}{r}\int^r_s dr' r'^2 \rho_{\text{NFW}}(r')} \nonumber \\ &=& \color{black}
    \sqrt{4\pi G\rho_0r_s^2}\ 
    \sqrt{\frac{1}{x}\left[\ln{|1+x|}-\frac{x}{1+x}\right]}\ .
    \label{eq:vrot}
 \end{eqnarray}
\textcolor{black}{This function of $x$ is universal: once the position of the maximum has been fixed by $r_s$, the physical parameters then only affect the height of the curve, but not its shape, entirely controlled by $r/r_s$. This means that \emph{every spherical NFW halo} needs to see a decreasing velocity after its maximum at $x\simeq 1.718$ as given in table~\ref{tab:fallNFW}}.
\begin{table}[h!]
\caption{\color{black} Values of the universal velocity-function shape for a Navarro-Frenk-White profile as per Eq.~(\ref{eq:vrot}) starting at its maximum. For values of $r$ beyond the maximum of $V$, certainly by three times that $r_s$, the fall-off should be seen in the data (an equivalent statement holds for other profiles). It is the purpose of this article to see whether this is already the case across the database (at the present time, it is not). \label{tab:fallNFW}  }
\begin{center}
\begin{tabular}{|cc|cc|cc|cc|}
\hline 
$x=r/r_s$ & $V(r)/V(r_s)$ & $x=r/r_s$& $V(r)/V(r_s)$  & $x=r/r_s$& $V(r)/V(r_s)$& $x=r/r_s$& $V(r)/V(r_s)$\\ \hline
1 & 1 & 2 & 0.955 & 3 & 0.896 & 4 & 0.844 \\
5 & 0.800 & 6 & 0.764 & 7 & 0.732 & 8 & 0.704 \\
 \hline 
\end{tabular}
\end{center}
\end{table}

In Figure~\ref{fig:NFW} opening the article we observe the behaviour of that DM rotation curve of a NFW spherical halo using Eq.~\ref{eq:vrot}. \textcolor{black}{The $r$ and $V$ axes have physical units fixed} with parameters those of the Milky Way, to be specific, following~\citep{Fabrizio_Nesti_2013}, \textcolor{black}{but ignoring those units, the universal shape can be ascertained}.

The maximum velocity is found in that example plot (many similar ones will follow) \textcolor{black}{at a radius near 30 kpc, and for larger radii this model $V(r)$ function shows a slight declining behavior: at 60 kpc the DM rotation curve should have decreased a 5\% with respect to the maximum,   at 90 an 11\%,}%end color
starting to show the $V(r)\propto 1/\sqrt{r}$ characteristic of Keplerian orbits because most of the halo mass is already inside the orbit. 

On the contrary, many galaxies show the behaviour found by Rubin and collaborators, as exemplified on the right panel of the same Figure~\ref{fig:NFW}: no such tapering is visible in the velocity function. \textcolor{black}{This lack of confirmation by the data is of course not exclusive of the NFW model, but of any similarly convergent spherical mass distribution, which eventually needs to turn and follow a Keplerian profile.}

Several galaxies in the database can be well fit with the NFW model~\citep{Bariego_Quintana_2023}. 
At the same time, as plot in Figure~\ref{fig:NFW}, there are also many galaxies which look asymptotically flatter than NFW allows, and its parameter freedom seems insufficient. 

Indeed, the only role of the density parameter $\rho_0$ is to set the vertical scale (maximum velocity) of the curve in the combination  $\sqrt{\rho_0 r_s^2}$. 
The shape of the rotation curve is then determined by one parameter alone, $r_s$, which sets the scale of the OX axis. As we plot the data in physical units, kiloparsec, there is a continuum of possible rotation curves, but mind that they are all obtained from the same universal shape by dilatations or contractions along that OX (radial variable) axis. Then, if we make the curve asymptotically flatter to match the observational data, the initial slope for small r is also weaker. Both a steeply growing curve at small r and a slowly declining one (or not declining at all) at large r cannot be obtained from NFW alone. \textcolor{black}{This is observed in many rotation curves even after subtracting the visible-matter contribution.}

\textcolor{black}{
Finally, we do not want to focus the discussion on a particular model or parametrization; any mass-convergent DM distribution eventually must show turning galactic rotation curves, as possibly seen in the Milky Way.
Thus we have framed the work in terms of a model-independent approach to data, which is why concentrate on asking the question, ``is the final slope consistent or not with zero?”, in a purely data-driven way. 
Then, we attempt an investigation of the SPARC data base to ascertain whether the turnover can be statistically asserted or that the data base can not distinguish from flat rotation curves. }%end document

Because the baryonic-mass distribution in the Milky Way is known to distances of $\sim 50$ kpc, and is well confined, in the external regions of the galaxy the centripetal force contributed by baryons is typically subdominant and returns the expected velocity decreasing as $r^{-1/2}$; therefore, we do not expect the baryonic component of Milky-Way like galaxies contributing significantly to the observed flattening, unnatural for a spherical NFW profile~\footnote{
The only spherically-symmetric viable mass distribution in those outer galactic confines is the power-law $\rho\propto r^{-2}$,
fine-tuned to precisely that exponent, only explainable if one accepts a thermalized halo, another questionable assumption given that the microscopic DM properties are unknown. 
Alternatively, one can opt to accept prolate or even cylindrical sources of gravity, which would make the constant $V(r)$ natural outside the DM distribution~\citep{Llanes_Estrada_2021, Bariego_Quintana_2023, Bariego_Quintana_2024}.}.

Pushing to still larger distances, a recent study extended the would-be rotation curve of four galaxies to distances of hundreds of kpc and even a Mpc using weak gravitational lensing~\citep{2024ApJ...969L...3M}. 
That work concluded that the rotation curves at those extreme distances, if they could be directly measured, would continue to show constant $V(r)$ rotation velocity, unlike a spherical halo with a Navarro-Frenk-White density profile~\citep{1985}. 
Other authors, concentrating on our own Milky Way galaxy, show evidence in favour of a possibly important Keplerian decline in its rotation curve \citep{2019ApJ...871..120E, Jiao_2023,klačcka2025decreasemilkywayrotation} but it is still unclear how much is baryon-driven. 
The remainder, the part of the rotation velocity attributed to the supposed DM, as shown in section~\ref{subsec:MW} is less settled, with disparate slopes shown in table~\ref{tab:MW}.

Our aim is to ascertain whether the further-most part of the rotation curve starts decreasing or not, that is, a test only on the final derivative $V'(r)_{\rm rmax}$, by statistical analysis of the SPARC-database~\citep{2016AJ....152..157L}. There, galactic rotation curves measured from the Doppler Effect in the H\textsc{i} and H$\alpha$ lines for 175 spiral and irregular galaxies are provided. Additionally, the baryonic component rotation curves for the gas, bulge and disk of each are estimated based on surface photometry of galaxies at $3.6$ $\mu$m. The baryonic rotation curve is the result of the squared sum of all:
\begin{equation}
    v_{\text{BA}} = \sqrt{\Upsilon_{\text{bulge}} \cdot v_{\text{bulge}}^2+v_{\text{gas}}^2+\Upsilon_{\text{disk}} \cdot v_{\text{disk}}^2}. 
    \label{eq.2}
\end{equation}
The criterion adopted by the SPARC collaboration \citep{2016AJ....152..157L} for the mass-to-light ratios $\Upsilon$ at 3.6 $\mu$m is $\Upsilon_{\text{bulge}}=1.4\Upsilon_{\text{disk}}$
and $\Upsilon_{\text{disk}}=0.5$ $M_\odot/L_\odot$, where $M_\odot/L_\odot$ is the mass-to-light ratio for the Sun. These values are induced from stellar population synthesis (SPS) models, and provide the best fit to the Tully-Fisher relation \citep{2014AJ....148...77M, 1977A&A....54..661T}. We adopt at face value their extracted visible-matter distributions.

When accounting for a pressumed DM halo, there is an additional contribution to the total rotation curve, $v_{DM}$, 
%which is also added in quadrature (because the corresponding centripetal force is added linearly and $F_c\propto v^2$), 
\begin{equation}
    V = \sqrt{V_{DM}^2+V_{BA}^2}.
    \label{eq.1}
\end{equation}

In Section \ref{sec:methodology} the set of galaxies is trimmed off irrelevant ones for the analysis (mostly because of the measurements not extending far enough from the galactic center),
and  statistical tests are deployed to decide whether the \textcolor{black}{dark matter} slope $V'(r)|_{r=r_{\rm max}}$ is compatible with zero or noticeably negative \textcolor{black}{(as spherical halo models predict $V'(r)<0$ for $r$ large enough)}.
The quantity which we actually fit is $v_{DM}(r)$, obtained by quadratically subtracting  $v_{BA}(r)$  from $V(r)$.

%, using baryonic rotation curve obtained from observations for the luminous components and
%the total rotation curve $V(r)$ from the H\textsc{i} lines

First, in Subsec. \ref{sec:1} we evaluate the slope at the end of the galactic disk, a distance at which only small quantities of baryonic matter are expected.
This proceeds by fitting a linear function to the last points of the DM rotation curve. Then, in Subsec. \ref{sec:2}, we study $V'(r)$ to understand whether a general downsloping trend for the rotation velocity in galaxies is obeyed by the last points or whether $V(r)$  remains flat until very distant radii. In Subsec. \ref{sec:3} we select appropriate hypothesis tests to study the general trend preferred by the galactic rotation curves in the selected sample. The results of the analysis are then given in Section \ref{sec:results}, and we summarize in Section \ref{sec:conclusions}. \\

%%%%%%%%%%%%%%%%%%%%%%%%%%%%%%%%%%%%%%%%%%%%%%%%%%%%%%%%%5
\section{Methodology}
\label{sec:methodology}
%%%%%%%%%%%%%%%%%%%%%%%%%%%%%%%%%%%%%%%%%%%%%%%%%%%%%%%%%%

In this section we fit the last data points of the DM rotation curve $V(r)$ for each galaxy to ascertain whether their general trend  is a decrease or rather a constant value of $V$. 
The largest-$r$ points will be fit to a linear function  \begin{equation} \label{linearfit}
    V(r) = A\cdot r + B\ ,
\end{equation} using $\chi^2$ minimization. Then we perform a hypothesis test where our null hypothesis is the null slope $V'=A=0$ and the alternative hypothesis a negative slope $A<0$.

We accept the SPARC data for the total rotation curve and for the baryon-contribution separation at face value. In isolating this baryon contribution (to more clearly see DM), a choice had to be made about the numerical value of the light to mass ratio. The SPARC data table (see note 3 in its header) is rendered with the baryonic disk and bulge at 
$M/L=1\ M_\odot/L_\odot$ , but the refereed SPARC publication \citep{2016AJ....152..157L} shows the sensitivity to other choices as well as detailing the careful treatment and separation of the baryonic matter contribution. Looking at the galaxy plots which we have produced, we do not see how tuning this parameter to another value would alter our conclusions.

%%%%%%%%%%%%%%%%%%%%%%%%%%%%%%%%%%%%%%%%%%%
\subsection{Selection of and data points for each of them}
\label{sec:1}
%%%%%%%%%%%%%%%%%%%%%%%%%%%%%%%%%%%%%%%%%%%

Broadly speaking, we see three  rotation-curve types in the SPARC data, with the behavior of $V$ as a function of $r$ being:
\begin{itemize}
    \item increase - constant - decrease
    \item increase - constant
    \item increase\ .
\end{itemize}
To ascertain whether there is or not a statistically significant decrease, we cannot accept galaxies that still show decidedly positively sloped $V$. We should only accept galaxies measured outside of the increasing part: in spirals we would observe a constant $V$ (maybe decreasing at some point?) and in dwarf galaxies maybe we would observe a fall-off. The question is whether the slope in this set does turn negative or not.

The first (and simplest to understand) requirement to be met is that the total rotation curve, in the largest-$r$ data points,  not be dominated by the baryonic matter of the galaxy. That is, the baryonic rotation curve from Eq.~(\ref{eq.2}) must then be less significant than the $V_{DM}$ that we infer from Eq.~(\ref{eq.1}). This condition avoids difficult behaviour resulting from the internal galactic structure, where the 
%\textcolor{black}{(shape-richer, depending on its extraction)}
baryonic contribution to the rotation curve can be larger than the DM one, and keeping such galaxies or such points within a galaxy would unduly distort the information about DM extracted.

The second condition is less straightforward.
One could think about setting a cut at a chosen physical radius, but each rotation curve has a different morphology (due to the underlying, light or dark, matter distribution). Such setting  would not be robust enough.
We wish to avoid human selection bias in deciding which points are to be included in each rotation curve, since the galaxies in the database are of different sizes and sampled at different $r$ intervals. We thus need an automated selection criterion for the last data points in each (total) rotation curve which can be applied to all galaxies without human intervention.

Therefore we implement these two simple, data-driven conditions:
first, being outside of the increasing part of the rotation curve
$V_{\rm BA} < V_{\rm DM}$  taking at face value the separation of the SPARC collaboration. This condition gets rid of rotation curves  that are only seen to increase, probably including most of the dwarf galaxies.

Then, for each galaxy in the SPARC data set with $V(r)$ sampled at $n$ data points, we select all possible subsets of four consecutive data points: $[S_1, S_2, ...]=[[V_{i+1}, ...,V_{i+4}],...]$ for $i=0,n-4$. We fit each subset $S_j$ to the linear function of Eq.~(\ref{linearfit}) with two free parameters $\{A, B\}$ 
using a $\chi^2$ function that we minimise:
\begin{equation}
    \chi^2 = \frac{(V_{theoretical}-V_{data})^2}{\sigma ^2_{data}(V)}
    \label{eq:chisq}
\end{equation}
where $v_{theoretical}$ is the linear parametrization (fittable function), and the total rotation curve velocity $v_{data}$ and the uncertainty $\sigma (v)$ are taken from the database. The $\chi^2$ minimization is performed by the least square method on each subset $S_1, S_2, ...$, proceeding from left (lowest $r$) to right (highest $r$) until the slope $A$ satisfies the  condition 
\begin{eqnarray}\label{StartCondition} 
A \leq A_{\rm max} = 0.9  \ \ \rm km/(s\ kpc)\ .
\end{eqnarray}
Given that typical galaxies have velocities of order of magnitude 100 km/s and size tens of kiloparsecs, when we reach this slope level (ten times smaller than the typical average ones over the entire range) we reasonably know that we are near the maximum of the $V(r)$ curve if there is one. The criterion is uniformly applied to the entire database to reduce data selection bias. 
The numerical sensitivity to the particular choice of $A_{\rm max}$ in Eq.~(\ref{StartCondition}) will be given below in Subsection~\ref{subsec:sensitivity}.

Once the slope of $V(r)$ has decreased to the level of Eq.~(\ref{StartCondition}), we discard the data points that belong to subsets with smaller $i$ indices (thus having smaller $r$), namely  $\bigcup S_{j\leq i}=[V_0, ...,V_{i}]$ and keep the rest,  $\bigcup S_f=S_{total}-{\bigcup S_j}=[V_{i+1}, \dots V_{n}]$. As an example, in Figure \ref{fig:NFW} the kept data points are shown in red.
 This procedure should ensure a reasonably consistent selection of the ``final'' few points of each rotation curve across the galaxy database. After the velocity slope decreases to this low value, we should reasonably be able to discern whether it keeps sliding below zero and becomes negative (Keplerian taper) or whether it remains consistent with zero, and learn about the asymptotic trend of velocity curves. 

It is important to note that in imposing these two conditions, (1) DM dominated rotation and (2) decreased slope, we will  have discarded a good number of galaxies; by the second condition those in which the rotation curve is monotonously increasing and never becomes constant or decreasing, indicating that the data was taken still deeply inside the matter/DM distribution.\\

%%%%%%%%%%%%%%%%%%%%%%%%%%%%%%
\subsection{Fitting method}
\label{sec:2}
%%%%%%%%%%%%%%%%%%%%%%%%%%%%%%
At this point we are left with a new set of data points $S_f=[V_{i+1}, V_{n}]$ for each of the remaining galaxies. The next step will be to make a fit of these fragments of rotation curves, but this time we consider the DM component instead of the total rotation curve.
\textcolor{black}{
The motivation for keeping only $V_{DM}(r)$ is clear, as we are intent on learning on the halo properties, particularly whether the imprint of ``edges" of the dark matter distribution can be seen in the data.}

We use Eq.~(\ref{eq.1}), the baryonic component $V_{BA}$ is taken from Eq.~(\ref{eq.2}) and the DM contribution will be fitted to a linear function of the type of Eq.~(\ref{linearfit}),
performing the least square minimisation method of the $\chi^2$ function in Eq.~(\ref{eq:chisq}).
%\begin{equation}
%    \chi^2 = \frac{(v_{theoretical}-v_{data})^2}{\sigma ^2_{data}(v)}.
%\end{equation}
We aim  at optimising the slope of each $V(r)$ and at understanding its behaviour in every galaxy of this data set.

%%%%%%%%%%%%%%%%%%%%%%%%%%%%%%%%%%%%
\subsection{Hypothesis testing}
\label{sec:3}
%%%%%%%%%%%%%%%%%%%%%%%%%%%%%%%%%%%%
\label{subsec:hypotest}
To probe the distribution of the fitted $V'(r)$ slopes over the galaxy sample, $[A_1, A_2, ..., A_{N}]$,  we will employ two statistical tests with two hypotheses each, as follows:
$$\begin{array}{cccc}
         & \text{Parametric: Z test }             & &\text{Non-parametric,}\\
         & \text{assuming a normal distribution}  & &\text{agnostic about distribution}\\
   H_0:  &A\geq0 & &A=0 \\
   H_1:  &A<0  & &A<0 
\end{array}
$$
First, we perform a parametric\footnote{We use ``parametric'' and ``nonparametric'' in the sense of generic statistical analysis, not, as sometimes seen, in the sense of whether rings (radial bins) at different values of $r$ are used to extract $V(r)$ nonparametrically or whether specific functions with free parameters are fit to the data.} statistical hypothesis test, the \textbf{Z-test}, to determine whether the \textbf{mean} across the galaxy sample of the estimated slopes, $\bar{A}=\sum_i A_i/N$,  is or not significatively lower than $\bar{A}=0$. This parametric test assumes a normal distribution for the slopes $A_i$. The test statistic $Z=\bar{A}/\sigma'(A)$ is calculated from the mean and its standard uncertainty $\sigma'(A)=\sigma(A)/\sqrt{N}$, the ratio of the standard deviation $\sigma(A)$ to the square-rooted sample size $N$. To determine the level of significance we compare the probability ($p$-value) of observing the mean $\bar{A}$ under the hypothesis $H_0$, that there is no tapering,  to a level of significance $\alpha$, and reject the null hypothesis when $p<\alpha$ which would establish the Keplerian behaviour.

Because it is \emph{a priori } not known whether the galaxy-end slopes $A=V'(r)$ are normally distributed, we need a second diagnostic. The null hypothesis of such second test will be to assume the \textbf{median} of all slopes to be zero, that is, it 
asserts that each individual-galaxy slope  has either sign with equal 50 $\%$ probability 
(independently of its absolute value).
The alternative hypothesis $H_1$, on the contrary, states that the $V(r)$ curves fit to data decrease, that is, in the alternative hypothesis the majority of data will fall on the negative side where $A=V'(r)<0$, as within a halo where the density has dropped so that little additional mass is enclosed. %as within an NFW halo. 

To contrast these two hypotheses we have chosen a natural non-parametric test,  the so-called \textbf{sign test}. It evaluates the symmetry of the distribution around $A=0$.

We calculate the number of values $N_+$ in our slope set  $[A_1, A_2, ..., A_k]$ that are positive, and also the number of those below 0, denoted in turn as $N_-$. Our fitted-slope data set will  approximately follow a binomial distribution, where each data point has a probability $q$ to be below zero; the total number of galaxies with negative terminal slope then follows a binomial distribution $N_-\propto \text{Binomial}(N, q)$. 

If the null hypothesis is true, we are to accept a probability of $q=50$ \% for any one galaxy to have either increasing or decreasing end velocity, and we expect that approximately half of the slopes are negative (as we have set the median value to be tested at $A=V'(r)=0$, Rubin's flat rotational curves). If the probability is statistically different from $1/2$, the data will reject the null hypothesis and possibly favour declining behaviour.
 %possibly favour NFW-like behaviour. 

We test for significance by computing the $p$-value for the binomial distribution. 
When the $p$-value is lower than a critical value, here chosen as 0.05 ($95$ $\%$ confidence) we say that there is enough evidence to conclude that the median is below 0, reject the null hypothesis $H_0$ and  accept the alternative hypothesis $H_1$.

However, if the opposite were true, and the $p$-value larger than 0.05, we could not reject the null hypothesis of flat rotation because of scant evidence to confirm that the median of the slopes is lower than 0. Strictly speaking, this would not prove $H_0$ to be true either, there would just not be enough evidence to discard it. 

The result of these two tests will be given presently in Section~\ref{sec:results}.

But before, we propose a second non-parametric test to contrast both hypotheses: the \textbf{Wilcoxon test}. This test is designed to look for evidence on whether the median of the distribution of the galactic-slopes set is significantly different from zero, irrespective of whether larger (increasing $V(r)$) or smaller (decreasing $V(r)$) than 0. The test does not assume a normal Gaussian distribution from which the data is sampled, but a symmetric distribution with equal tails to the sides of its median (which the sign test does not, of course). The null hypothesis is here that the data for $A$ is symmetric around $0$ (Rubin value). The one-sided alternative hypothesis is then taken as the data being symmetric respect to some median value which is negative (declining behaviour). 

%%%%%%%%%%%%%%%%%%%%%%%%%%%%%%%%%%%%%%%%%%%%%%
\section{Results}
\label{sec:results}
%%%%%%%%%%%%%%%%%%%%%%%%%%%%%%%%%%%%%%%%%%%%%%

%%%%%%%%%%%%%%%%%%%%%%%%%%%%%%
\subsection{Sensitivity to the value of $A_{\rm max}$ from Eq.~(\ref{StartCondition})}
\label{subsec:sensitivity}
%%%%%%%%%%%%%%%%%%%%%%%%%%%%%%
In the first order of business, let us discuss the sensitivity to the choice of $A$ of Section \ref{sec:1} in Eq.~(\ref{StartCondition}).
Because the selection of data points to be fit depends on that choice, we need to compare several values of this $A\leq A_{\rm max}$ condition, used to select galactic data points where the rotation curve starts bending towards either a constant or a negative slope, 
as opposed to the positive one typically displayed by the inner-most points of a galactic rotation curve.

 With this intention we test 10 values of $A_{\rm max}$ ranging between 0 and 1. Then, we rank the linear models represented by each of those values from best to worst in Table~\ref{tab:models}. For each galaxy we rank each of the 10 models by their $\chi^2/$dof value, to then compute the average rank (Avr rank) for each model across all galaxies.

 The lower the average-rank index, the better the performance of the model. In the table we see that most of the models have a very similar average rank, so we expect them to similarly describe the rotation data. 

 In order to understand the agreement among galaxies on the ranking which they assign to each of the values of $A_{\rm max}$, we deploy Kendall's W test. While the average rank informs about the mean position of each model across galaxies, Kendall's W reports the consistency in ordering the models across galaxies.
 
 The concordance coefficient $W=12\ S/(m\ n^3-m\ n)$ is calculated from  the number of galaxies that the 10 models have in common, $m=93$;  the number of models, $n=10$; and $S=\sum^n_{i=1}(R_i-\bar{R})^2$, the sum of the ranks $R_i$ and the mean of the ranks $\bar{R}$ for each model. The coefficient W takes values range between 0 (full disagreement) and 1 (full agreement).

We obtain a very low $W=0.011$, which means that there is little agreement in the models' rankings between galaxies: different values of $A_{\rm max}$ do better in describing different subsets of the galaxies, with no clear preference for any one.

Indeed, when many galaxies rank the values of $A_{\rm max}$ differently, the average rank can be close to a middle value (5 out of 10 in this case), but the rankings are not consistent across galaxies. All the values of the maximum slope at which we start considering that we enter the final tail of the rotation curve perform very similarly, \textcolor{black}{as seen in Table~\ref{tab:models}, we also observe similar p-values arising. Therefore, we decide that there is no preference over the selection of a $A_{\rm max}$ value}. %so beyond choosing a natural one, we  examine the $p$-value trend that arises from all of them, as seen in Table~\ref{tab:models}. 

\begin{table}[h!]
    \centering 
    \begin{tabular}{|c|c|c|c|c|c|c|c|}\hline  
        Ranking & $A_{\rm m ax}$ & Avg rank & binomial $p$-value & Wilcoxon $p$-value & normal $p$-value & \# of galaxies & $A<0$\\ \hline    
        1 & 0.0 & 4.87 & 1.0 & 1.0 & 0.9 & 90 & 22 \\ \hline  
        2 & 0.1 & 5.17 & 1.0 & 1.0 & 0.9 & 94 & 21 \\ \hline  
        3 & 0.2 & 5.34 & 1.0 & 1.0 & 0.9 & 97 & 21 \\ \hline  
        4 & 0.3 & 5.38 & 1.0 & 1.0 & 0.9 & 97 & 20 \\ \hline  
        5 & 0.4 & 5.47 & 1.0 & 1.0 & 0.9 & 97 & 20 \\ \hline  
        6 & 0.6 & 5.57 & 1.0 & 1.0 & 0.9 & 99 & 20 \\ \hline  
        7 & 0.7 & 5.69 & 1.0 & 1.0 & 0.9 & 103 & 20 \\ \hline  
        8 & 0.8 & 5.75 & 1.0 & 1.0 & 0.9 & 110 & 18 \\ \hline  
        9 & 0.9 & 5.86 & 1.0 & 1.0 & 0.9 & 112 & 16 \\ \hline  
        10 & 1.0 & 5.88 & 1.0 & 1.0 & 0.9 & 113 & 16 \\ \hline  
    \end{tabular}    \caption{Set of models characterized by $A_{\rm max}$ (in km/(s\ kpc)) ordered from best to worse; their average rank (according to the 93 galaxies that the models have in common); their $p$-values for different statistical tests (non-parametric sign \& Wilcoxon tests, and a parametric Z-test); the number of galaxies which remain in each model and the number of galaxies with a decreasing slope $A<0$ \textcolor{black}{(close to about a quarter of the total)}. }
    \label{tab:models}
\end{table}

The conclusion of the computation in that table is that choosing $A_{\rm max}$ to be 0.9 is not substantially different from selecting basically any number between 0.5 and 1.0, so it merits no further discussion and we proceed to assess the distribution of slopes. 

\newpage
%%%%%%%%%%%%%%%%%%%%%%%%%%%%%%%%%%%%%%%%%%
\subsection{Rotation-velocity radial dependence at large distances}
%%%%%%%%%%%%%%%%%%%%%%%%%%%%%%%%%%%%%%%%%%

\textcolor{black}{In the previous sections we made a fit of the DM rotation curve of the SPARC galaxy data set to evaluate the general trend at the end of the galactic disk.}  After having decided which ones are the far-out data points in each rotation curve as per Section \ref{sec:1}, the sample size is smaller:  about 4/7 of the original 175 rotation curves from the SPARC database pass the cut (see Table~\ref{tab:models}) leaving out those which have sizeable positive slope basically to the end of the measured rotation curve. 
Taking those galaxies out somewhat biases the analysis in favour of  \textcolor{black}{a declining rotation curves} instead of Rubin's behaviour (flat rotation curve). This happens because the number of galaxies with $A>0$ is decreased, which is why it appears that selecting the somewhat higher $A_{\rm max}=0.9$ is a better compromise than choosing a smaller value thereof.
This is because those values of $A_{\rm max}$ in Table \ref{tab:models} which are closer to zero will typically produce data sets where the general trend of the rotation curve will be to either decrease or stay constant, with a smaller chance to be slightly increasing, but due to a selection bias, so we keep the looser cut. 
Note that if any systematic bias would arise from eliminating the clearly rising rotation curves, it would favour the negative-derivative hypothesis: but our analysis fails to discard the null one, so including those galaxies would only improve the statistical significance of the test result, in  favour of our conclusion.

In~\ref{sec:appendix}) a few sample galaxies which do pass that cut are shown; those in figure~\ref{fig:examples1} have slightly positive velocity slope, those in~\ref{fig:examples2} rather negative, but many of them compatible with zero within one standard deviation anyway. Plots of the rotation curves of \emph{all} galaxies in the SPARC database and the fitted slopes when appropriate are left for the supplementary material.

\begin{figure}[h!]
    \centering
\includegraphics[width=0.5\linewidth]{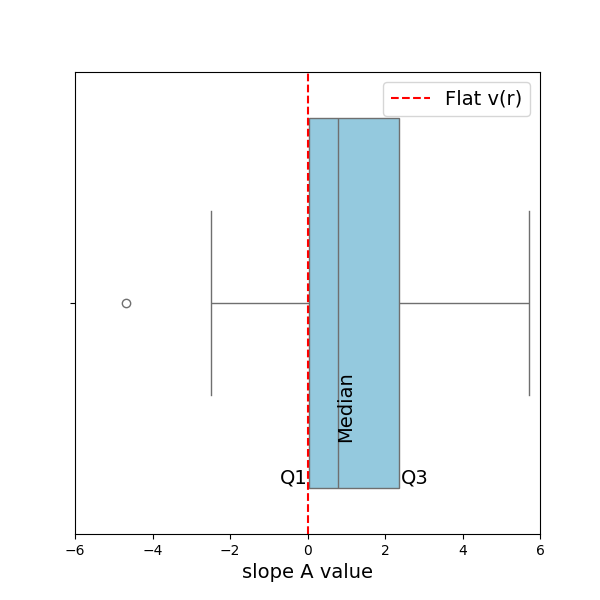}
    \caption{The distribution of terminal slopes $V'(r)$ is not symmetrically distributed around $A=0$, and does not seem to have a preference for negative slope values.
    The black box represents the \textbf{interquartile range  (IQR)} for the distribution of the slopes fit to the rotation data, from the first (Q1) to the third (Q3) quartiles, and they are comprised within the small range of 2 km/(s\ kpc) (units of the OX axis).  The black line inside that box lies at the median of the $V'(r)$ distribution and is not far from 0. The whiskers extend to 1.5$\times$IQR above and below Q1 and Q3, and outside these values we find the occasional outlier (empty circle). }
    \label{fig:distribution1}
\end{figure}

Next, as explained in Section \ref{sec:2}, a $V'(r)$ end derivative is extracted for each of the remaining galaxies, after fitting to Eq. \ref{linearfit}. In Figure~\ref{fig:distribution1} we represent  the distribution of the data in a box plot that extends from the first to the third quantiles:  SPARC rotation curves demonstrate a preference for rather positive, increasing slopes, than negative ones, but the median is very close to zero, and the interquartile range is narrow, with a width of about \textcolor{black}{$2$ km$\cdot$s$^{-1}$kpc$^{-1}$}, when the natural slope size (looking at the empirical rotation curves) is a much larger $10$ \textcolor{black}{km$\cdot$s$^{-1}$kpc$^{-1}$}. This means that the DM contributions to the rotation curves become, statistically speaking, rather flat at the end points.
\\

\begin{figure}[h!]
    \centering
\includegraphics[width=0.5\linewidth]{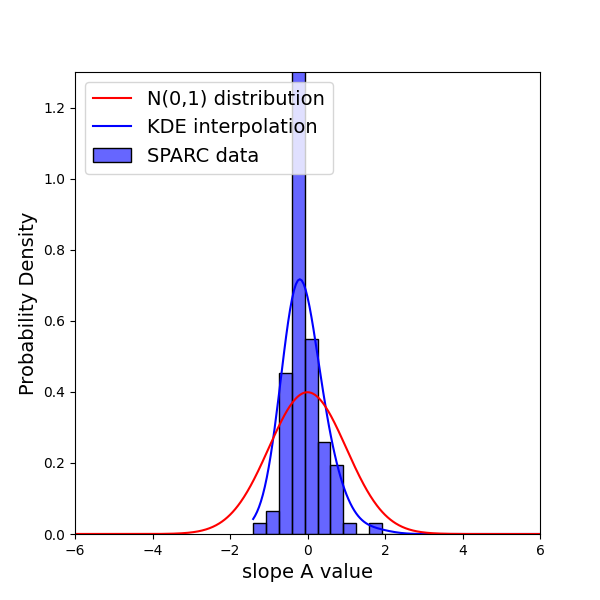}
\includegraphics[width=0.45\linewidth]{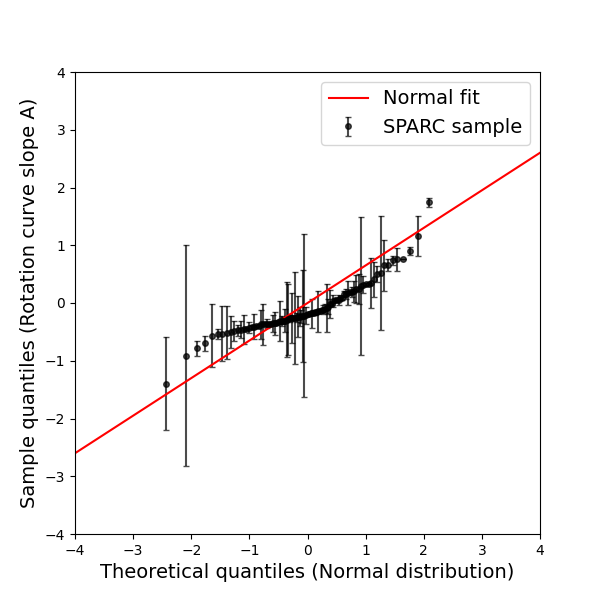}
    \caption{\textbf{Left:} 
    Histogram representation of the galactic end slopes together with its smoothing obtained by a Kernel Density Estimation (KDE, blue line) and compared to a normal distribution (red line). The symmetric shape centered near 0, as the statistics tests suggest, is clearly visible by eyeball. 
    \textbf{Right:} Quantile-quantile representation to assess whether the slopes are normally distributed as a Gaussian. The data points are sorted and represented against theoretical quantiles from a normal distribution. the red line is the best fit through the points. When the data points are close to the line the data are approximately normal, and this is reasonably the case, lending credibility to the parametric $Z$ test of Subsection~\ref{subsec:hypotest}.
    \label{fig:distribution2}}
\end{figure}

Then in the left panel of Figure~\ref{fig:distribution2} we give the distribution of the fitted SPARC-sample slopes in histogram form, and again see that it is rather peaked around zero.  This is corroborated by the smoothed Kernel Density Estimator of the histogram, which is rather well centered at zero. We also plot a Gaussian centered at zero with a deviation of 1 \textcolor{black}{ km$\cdot$s$^{-1}$kpc$^{-1}$,   the units of the $V(r)$ slope}.
\begin{equation}
    N(\mu=0,\sigma^2=1)= \frac{1}{\sqrt{2\pi}} e^{-x^2/2}
\end{equation}
which is seen to be a bit more fat-tailed and less peaked at the center, but close enough to discuss the normal variance of the data.
The quality of the Normal-Distribution approximation to the data is given in the corresponding right plot, the quantile-quantile representation, which confirms that the Gaussian approximation is reasonable and one can discuss the distribution in terms of a mean and standard deviation, as well as run a parametric test, the $Z$-test as described in Subsection~\ref{subsec:hypotest}. 

This is why we report both parametric (accepting a normal distribution of slopes) and non-parametric (for a sample that does not necessarily follow a normal distribution)  hypothesis tests to asses the general trend of the velocity slope sample selected off the SPARC database.

We showed the resulting $p$-values for the tests on the slope \textcolor{black}{which are quite similar}, in addition to the sensitivity to the different choices of $A_{\rm max}$, in Table \ref{tab:models}. That includes the Sign, the Wilcoxon and the Z tests already described. 
In that same table we find the number of galaxies that display a slightly declining DM rotation curve. The $p$-value for each $A_{\rm max}$ indicates the position of the sample within the test statistic distribution. A low $p$-value would suggest that the sample lies in the tail of the distribution,
increasing the odds of rejecting the null hypothesis in favour of the alternative ($H_1$). A higher $p$-value places the sample near the center of the test statistic distribution, favoring $H_0$. In our analysis, all values of $A_{\rm max}$ are conducive to large $p$ values, \textcolor{black}{providing no evidence against the null hypothesis with negligible sensitivity to $A_{\rm max}$}. 

The high values of $p$ obtained indicate that the data aligns with the center of their test statistic distributions, rejecting $H_1$: the overall behaviour of the database is rather consistent with rotation curves being flat as far as the measurements extend.

%To be specific, one can quantify how large these $p$ values are by comparing them to conventional significance thresholds $\alpha = 0.05$ (95\% significance) and $0.01$ (99\% significance). 
Since the obtained $p$  \textcolor{black}{$\sim$ 0.9--1} are way larger than both 0.05 (95\% significance)  and 0.01 (99\% significance), we fail to reject the null hypothesis ($H_0$) with any confidence.
As a result, there is no statistical evidence to defend a decreasing slope for large $r$, and the galactic rotation curves lend no support to the \textcolor{black}{convergent-mass-halo picture at large $r$ (yet?)}. In this analysis we very clearly fail to reject $H_0$, as SPARC data provides no support for the alternative hypothesis $H_1$ -- decreasing behaviour in the final points of the rotation curve--, over the $H_0$ hypothesis, namely  a constant $V(r)$.

%%%%%%%%%%%%%%%%%%%%%%%%%%%%%%%%%%%%%%%%%%%%%%%%%%%%%%%%%%%%%%%%%%%%%%%%%%%%%%%%%%%%%%%
\subsection{Testing Keplerian decline with Milky Way data sets}\label{subsec:MW}
%%%%%%%%%%%%%%%%%%%%%%%%%%%%%%%%%%%%%%%%%%%%%%%%%%%%%%%%%%%%%%%%%%%%%%%%%%%%%%%%%%%%%%%

\begin{figure}[h!]
    \centering
    \includegraphics[width=0.8\linewidth]{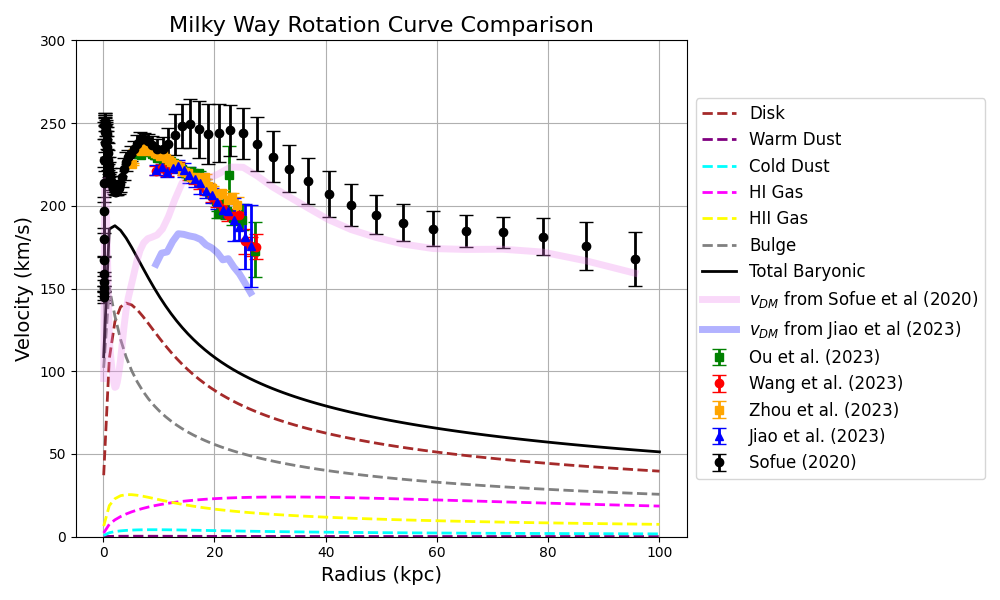}
    \caption{Milky Way's total rotation curve as compiled by \citep{Sofue:2020rnl} (the uppermost data set with $V(r)$ not yet falling), as well as the latest measurements reported by \citep{ou, wang, Zhou_2023, Jiao_2023}. These data show a steady decline of the rotation curve for $r>15$ kpc. The lines underneath show the contribution of ordinary baryonic matter to the rotation curves, modelling the galaxy components: disk, bulge, gas and dust. At these distances, baryonic contributions are still important. The inferred DM contribution is dominant in Sofue's compilation at large distance and does seem to flatten out.    \label{fig:MWtest}}
\end{figure}

A salient galaxy which is not in the SPARC database but can be analysed with modern observatories such as GAIA is our own Milky Way. The near-range measurements make them perhaps more reliable; at the same time, the modelling of the baryonic contributions is also more constrained.

In this subsection we apply the same methods used on the galaxies in the SPARC database to different contemporary measurements  for our Galaxy. In Figure \ref{fig:MWtest} we represent the latest measured rotation curves (\citep{ou}, \citep{wang}, \citep{Zhou_2023}, \citep{Jiao_2023}) with the compilation of older measurements from~\citep{Sofue:2020rnl}.  The newest data sets point out to a Keplerian decline in the total rotation curve.

In this subsection we cursorily study this data and separate the DM rotation contribution.  For this purpose we accepted the model of the baryonic components of~\citep{ou}, where the bulge is described by the Hernquist potential and the rest of the components (disk, gas and dust) are modelled by a double exponential density profile. The numerical values necessary for this baryonic modelling are directly taken from 
Table 2 of~\citep{ou}. 

We evaluate the linear behaviour of the DM rotation curves in three intervals: one  starting at $5$ kpc (to avoid contributions to the motion from the galactic bar) until $15$ kpc, the second region between 10 and 20 kpc (where the 2023 sets which benefit from Gaia data have good overlap)
and the third, outer one starting at $15$ kpc (where a steeper decline is suggested by the newer analyses).
We repeat the fitting procedure described in Section \ref{sec:2} with a two degrees of freedom linear function; the fits to the DM rotation curves are presented in Figure \ref{fig:appendix_MW}. In Table \ref{tab:MW} we describe results of the slopes that we fit in each region, and we add the goodness of fit. We observe that in previous measurements of the total MW rotation curve the fall-off is not as pronounced as in the newer  measurements.
We have not bundled the MW data with that from the SPARC database as it has a different intellectual import and it is not clear what weight it should carry in the averaging. Should it be weighed equally to other galaxies in the database, it would have small impact on our overall statistical conclusion. We believe it is best to show it separately. 
\textcolor{black}{As the data stands, one would need to explain why a decline is not visible in the SPARC galaxy database but seems to be present in the Milky Way. 
A plausible reason is that the SPARC rotation curves do not yet reach far enough distances; this could be explored with MeerKAT's MHONGOOSE \citep{deBlok2024} and the future Square Kilometer Array, by identifying the velocity of gas clouds in outer galactic reaches with velocity precision down to km/s.
}

\begin{table}[h]
\centering
\begin{tabular}{|l|c|c|c|c|c|c|}
\hline
Source & Slope & $\chi^2/N_{dof}$ & Slope  & $\chi^2/N_{dof}$ &
Slope  & $\chi^2/N_{dof}$ \\
& $r\in [5,15] $ kpc & & $r\in[10,20]$ kpc & & $r>$15 kpc & \\
\hline\hline
\citep{Sofue:2020rnl} & $4.21 \pm 1.41$ & 0.15 & $2.09\pm 3.73$  &  0.08 &  $-0.69 \pm 0.14$ & 0.27 \\
\hline
\citep{ou} & $1.87 \pm 0.09$ & 4.87 & $0.37 \pm  0.09$ & 6.65 &  $-2.1 \pm 0.2$ & 3.44 \\
\hline
\citep{wang} & $3.73 \pm 0.72$ & 0.64 & $0.43 \pm  0.35$ & 2.13 & $-2.14 \pm 0.28$ & 0.85 \\
\hline
\citep{Zhou_2023} & $1.54 \pm 0.04$ & 28.05 & $0.32 \pm  0.05$ & 7.8 & $-0.54 \pm 0.17$ & 1.68 \\
\hline
\citep{Jiao_2023} & $3.86 \pm 1.07$ & 0.21 & $0.52 \pm  0.53$ & 0.89 & $-2.44 \pm 0.72$ & 0.06 \\
\hline
\hline
Combined (Ou+ & & & & & & \\ Zhou+Jiao+Wang) & $1.51 \pm 0.04$ & 12.52 & $0.28 \pm  0.04$ & 5.67 & $-1.57 \pm 0.11$  & 2.75   \\
\hline
\end{tabular}
\caption{ Fit of the DM contribution to the rotation curve  of the Milky Way depicted in Figure \ref{fig:MWtest}. Three regions are fit: $r=5-15$ $kpc$, an intermediate one, and $r>15$ $kpc$. Here we quote the numerically fit slope in both intervals and the goodness of the fit considering the number of degrees of freedom ($N_{gdl}=2$). The newer sets are compatible with each other and can be combined in one, shown in the last row. The separated DM contribution to $V(r)$ seems to be tapering off. The data of~\citep{Sofue:2020rnl} for $r>50$ kpc seems to flatten out (see figure~\ref{fig:MWtest}) beyond what is here apparent. }
\label{tab:MW}
\end{table}

\newpage

%%%%%%%%%%%%%%%%%%%%%%%%%%%%
\section{Summary and conclusions}
\label{sec:conclusions}
%%%%%%%%%%%%%%%%%%%%%%%%%%%%

In this analysis we have tested the behaviour of the final points in the rotation curves for the SPARC data set with the intention of discriminating between \emph{(i)}  a decreasing behaviour, typical of Keplerian orbits (and also expected in Navarro-Frenk-White and other spherical halo models for $r$ beyond the matter distribution) and \emph{(ii)} a steady, non-decreasing behaviour as described by V. Rubin. 

We have left out from the galaxy sample any of them  with an indefinitely increasing behaviour, checking the irrelevance of how to choose this data cut in Table~\ref{tab:models}. 
As we exclude galaxies that do not reach a maximum velocity or plateau, but continue rising indefinitely: and because the typical slope in the inner part of the galaxy is of order 10 \textcolor{black}{$kms^{-1}kpc^{-1}$}, we cut at 1 \textcolor{black}{$kms^{-1}kpc^{-1}$}   as a criterion for having reached a very small slope already. The larger this, the less stringent the cut. Note that $A_{\rm max}=1$ does not yield an absolutely flat $V(r)$, so one can possibly entertain the doubt about whether a tighter cut would result in, against our conclusion, an overall negative slope over the galaxy sample. Thus, we have studied values all the way to zero. 
As we observe from the Wilcoxon test, there is no preference over a particular value of $A_{\rm max}$ in the studied range.
On the upper end of this segment, loosening the cut for $A_{\rm max}>1$ includes more galaxies with positive slope and cannot change our statistical result, so it is a marginally interesting exercise and we do not consider values of $A_{\rm max}$ larger than 1.

\textcolor{black}{Our statistical analysis shows that the null hypothesis cannot be rejected, indicating that the rotation curves are consistent with being flat within the uncertainties. Any potential systematic bias introduced by excluding galaxies with clearly rising rotation curves would favour the negative-derivative hypothesis. Including these systems would therefore tend to strengthen, rather than weaken, the statistical support for our conclusion.}

To reduce the baryonic matter effect we have further issued a second analysis reducing the initial sample  by discarding any galaxies with an increasing baryonic rotation curve. The statistical test results, shown in Table~\ref{tab:modelsII}, are perfectly compatible with the earlier ones. 

For each galaxy that passes the cut we have selected the data points at large $r$ at far-out values where baryonic matter does not dominate or has little effect over the shape of the rotation curve. The objective has been to provide information useful for the behaviour of gravity (e.g. MOND-like theories) or to particle physics properties (as simulations depend on these, and they can output the halo shape and density profile) in a region dominated by DM, where baryonic matter is nearly absent. 
While at intermediate radial distances $V(r)$’s fall-off can be a complicated function due to galactic structure, at asymptotically large distances, if the matter distribution is as implied by cosmological simulations, the Keplerian behavior needs to be reached. This is why we concentrate on the last stretches of each rotation curve, to help decide whether the data show, or not, that asymptotic regime. \textcolor{black}{With current statistical analyses suggesting this regime has not yet been observed}

The final data points for each rotation curve have been fitted to a simple linear function with $\chi^2$  minimisation, comparing the theoretical predictions with the observed data. With the minimisation of the function we have obtained a sample of slopes, $V'(r)=A$, for the remaining galaxy sample. 
We have evaluated the general trend of the slope on the sample using parametric as well as non-parametric tests, to reach the conclusion that the null hypothesis is preferred within a 95\% of confidence. SPARC galaxies do not show a strong preference in favour of a decreasing trend in their rotation curve, but rather lean towards a constant rotation curve. In previous works (\citep{Bariego_Quintana_2023}) SPARC galaxies already demonstrated 
a preference for elongated shapes that naturally provide a flat rotation curve over spherical haloes with specific profiles which look isothermal for a small portion of the $r$ domain only, and which slightly decline later at larger $r$.

DM radii are often quoted to reach distances of order 1 Mpc, but this is at very significantly diminished DM density. Should they remained isothermal, with $\rho\propto 1/r^ 2$, the linear divergence in the total mass integral would quickly exceed the fraction of DM in the universe. 
Modern constraints on the halo mass function, {\it e.g.} \citep{2021A&A...645A.126C} typically
assume NFW-like profiles. If the haloes remain isothermal out to longer radii, they would probably create tension with the parameters of the halo-mass distribution function, which are constrained to order 10\% level, or the underlying models linking luminous to dark matter.

 Rotation curves in acceptable spherical DM haloes naturally decrease with power-laws near the Keplerian $V\propto r^{-1/2}$ which must manifest at sufficiently large distances from the galactic center, as seen in Figure \ref{fig:NFW}. 
 
 In outer galactic regions where baryonic matter becomes negligible, any decline in the DM rotation curve should directly manifest in the total rotation curve, providing a critical test for \textcolor{black}{standard profiles derived within DM structure formation} and collapse (and also probing modified gravity~\citep{Khelashvili:2024gus}
 ). The fact that most SPARC galaxies do not support this scenario raises an important question about the shape and distribution of DM haloes around galaxies, which has significant implications for DM searches and the nature of the DM components as reflected in their macroscopic, collective behaviour. Gravitational-lensing measurements far away from the galactic disks, as recently reported~\citep{2024ApJ...969L...3M}, might be required to understand the distribution of DM around galaxies where the would-be rotation curve seems to stay flat until distances where the concept of galaxy itself is questionable.

On the contrary, beyond this large SPARC dataset, new data for the $V(r)$ curve of our own galaxy has been recently reported (see Figure~\ref{fig:MWtest}), and it is seen to fall-off unlike the typical SPARC galaxy at analogous distances from the center. \textcolor{black}{Upcoming radio surveys could improve the situation by providing further reach in the variable $r$ for numerous galaxies.}

Given that our $p$-values exceed the significance level ($p > \alpha$), we conclude that the null hypothesis (flat rotation) adequately describes the database at 95\% confidence level (corresponding to approximately two standard deviations). Even if a stricter significance threshold were applied, the results would remain unchanged. \\ 

\textcolor{black}{Additionally,  
we have found that the DM slopes display a distribution compatible with a normal one around the slightly positive median (after cleaning out the clearly increasing $V(r)$ curves).
This can signify the accumulation of many small random fluctuations or measurement artifacts (via the central limit theorem), or it may also add a signal to the apparent connection between baryons and DM 
such as the Baryonic Tully-Fisher relation or the Radial Acceleration Relation~\citep{Li:2018}, if baryons are still playing a role in defining the rotation curve at the largest $r$ probed by SPARC.}

In summary, we cannot reject a final constant slope, $H_0$, because there is not enough evidence to support a decrease in the slope ($A< 0$) of $V(r)$. At the same time, we cannot confirm $H_1$ (Keplerian taper of $V(r)$), as the statistical analysis does not provide sufficient grounds to conclude that the slope decreases. Thus, while we do not definitely confirm either hypothesis, we establish that our data selection does not provide evidence in favor of $H_1$, and the decreasing behaviour in the final points of the rotation curve needed \textcolor{black}{standard spherical halo} models of galactic DM remains to be found.

\section*{Acknowledgments}
The authors thank clarifying exchanges with Jose Beltr\'an at the Univ. of Salamanca,  Tobias Mistele from the Frankfurt FIAS, Alexander Knebe from Univ. Autonoma de Madrid and Alfonso Lazo Pedrajas from Univ. of Valencia. Work partially supported by grant PID2022-137003NB-I00 of the Spanish
MCIN/AEI /10.13039/501100011033/
\\

Preprint issued as IPARCOS-UCM-25-039

\newpage

% Bibliography
\bibliographystyle{plainnat} % Style (e.g., plain, apalike, ieeetr)
\bibliography{references} % Name of your .bib file (without .bib)

\appendix

%%%%%%%%%%%%%%%%%%%%%%%%%%%%%%%%%%%%%%%%%%%%%%%%%%%%%%%%%%%%%%%%%%%%%%%%%%%%%%%%%%%%%%%
\section{ Statistical tests with a reduced galaxy sample}
%%%%%%%%%%%%%%%%%%%%%%%%%%%%%%%%%%%%%%%%%%%%%%%%%%%%%%%%%%%%%%%%%%%%%%%%%%%%%%%%%%%%%%%

A criticism which can be raised is a probable ambiguity in separating the respective contributions of baryonic and DM to the rotation-velocity curve, this is typically known as the disk-halo degeneracy.

For this reason, in this section we limit the analysis to a more restricted subset of the previously used SPARC data.
We here substantially tighten the data cut in subsection~\ref{sec:1} to discard those galaxies in which the baryonic rotation curve is not strictly decreasing all the way to maximum radius where each rotation curve ends.

The number of accepted galaxies (depending on the $A_{\rm max}$ cut described in subsection~\ref{sec:1}) remains at least 85 or more, which is still probably large enough to discuss the statistics (usually 30 samples are required when discussing normal distributions, or 15-20 on each side of a nonparametric test, as a rule of thumb). 

In Table \ref{tab:modelsII} we examine the $p$-value obtained for the subset of galaxies which have decreasing baryonic rotation curves at large radii. This alternative analysis is designed to reduce the effect of baryonic matter on the statistical distribution of the velocity curve at large radii.

\begin{table}[h!]
    \centering 
    \begin{tabular}{|c|c|c|c|c|c|c|c|}\hline  
        Ranking & $A_{\rm m ax}$ & Avg rank & binomial $p$-value & Wilcoxon $p$-value & normal $p$-value & \# of galaxies & $A<0$\\ \hline    
        1 & 0.0 & 4.79 & 1.0 & 1.0 & 0.99 & 86 & 20 \\ \hline  
        2 & 0.1 & 5.11 & 1.0 & 1.0 & 0.99 & 90 & 19 \\ \hline  
        3 & 0.2 & 5.28 & 1.0 & 1.0 & 0.99 & 93 & 19 \\ \hline  
        4 & 0.3 & 5.39 & 1.0 & 1.0 & 0.99 & 93 & 18 \\ \hline  
        5 & 0.4 & 5.48 & 1.0 & 1.0 & 0.99 & 93 & 18 \\ \hline  
        6 & 0.6 & 5.59 & 1.0 & 1.0 & 0.99 & 95 & 18 \\ \hline  
        7 & 0.7 & 5.71 & 1.0 & 1.0 & 0.99 & 98 & 17 \\ \hline  
        8 & 0.8 & 5.77 & 1.0 & 1.0 & 0.99 & 105 & 15 \\ \hline  
        9 & 0.9 & 5.89 & 1.0 & 1.0 & 0.99 & 105 & 13 \\ \hline  
        10 & 1.0 & 5.97 & 1.0 & 1.0 & 0.99 & 105 & 13 \\ \hline  
    \end{tabular}
    \caption{Same as table \ref{tab:models} but restricting the data to a subsample of galaxies displaying clearly decreasing baryonic rotation curves at large radii, to reduce the baryonic contamination to the DM contribution to $V(r)$.}
    \label{tab:modelsII}
\end{table}

The \textcolor{black}{$p$ values are very similar to the ones from table~\ref{tab:models}, in this case the null hypothesis can neither be discarded} and the Rubin-type of rotation curve must be declared compatible with the data.

%%%%%%%%%%%%%%%%%%%%%%%%%%%%%%%%%%%%%%%%%%%%%%%%%%
\section{Plots of a few rotation curves}
\label{sec:appendix}
\begin{figure}[h!]
    \centering
    \includegraphics[width=0.45\linewidth]{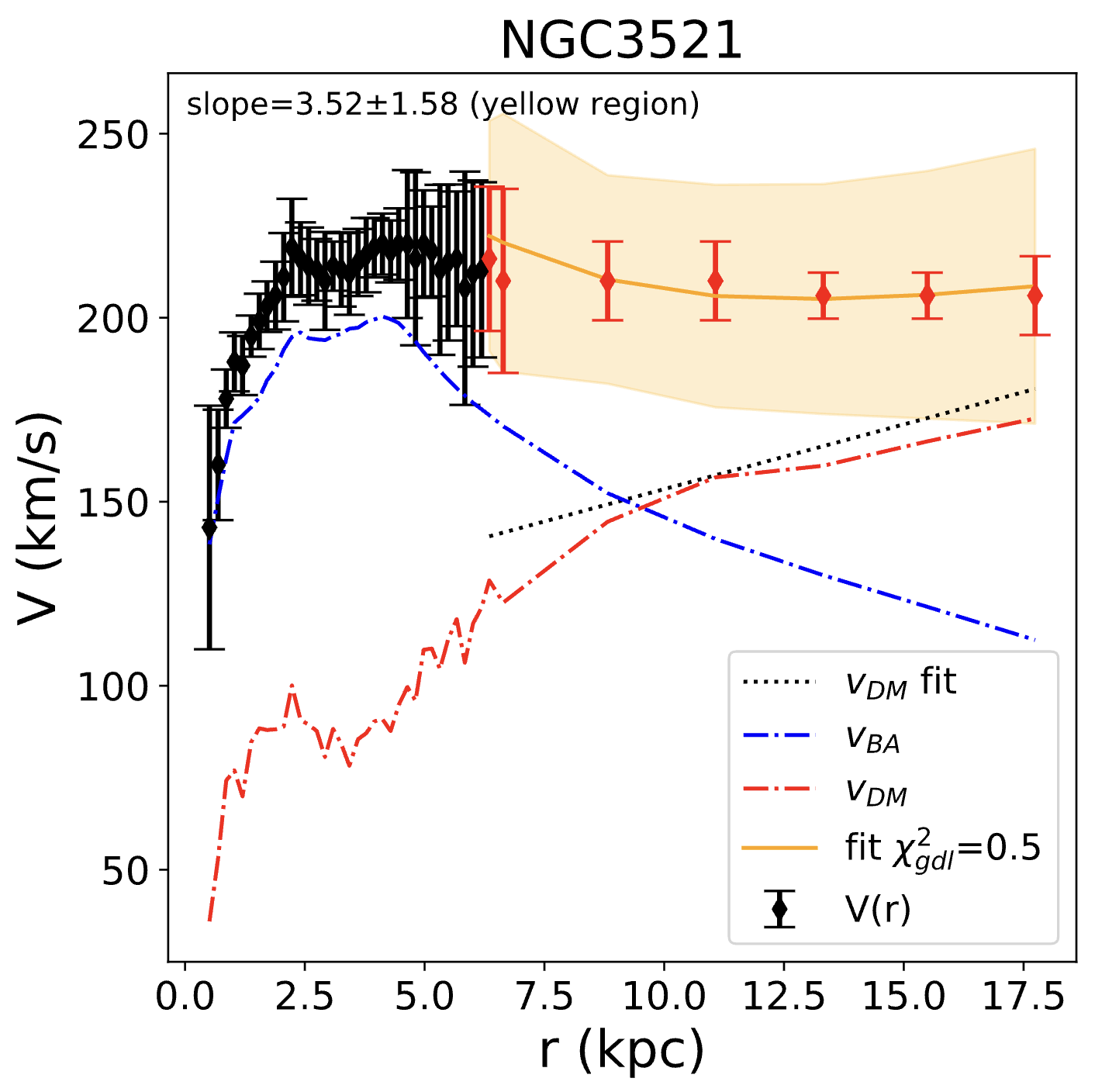}
    \includegraphics[width=0.45\linewidth]{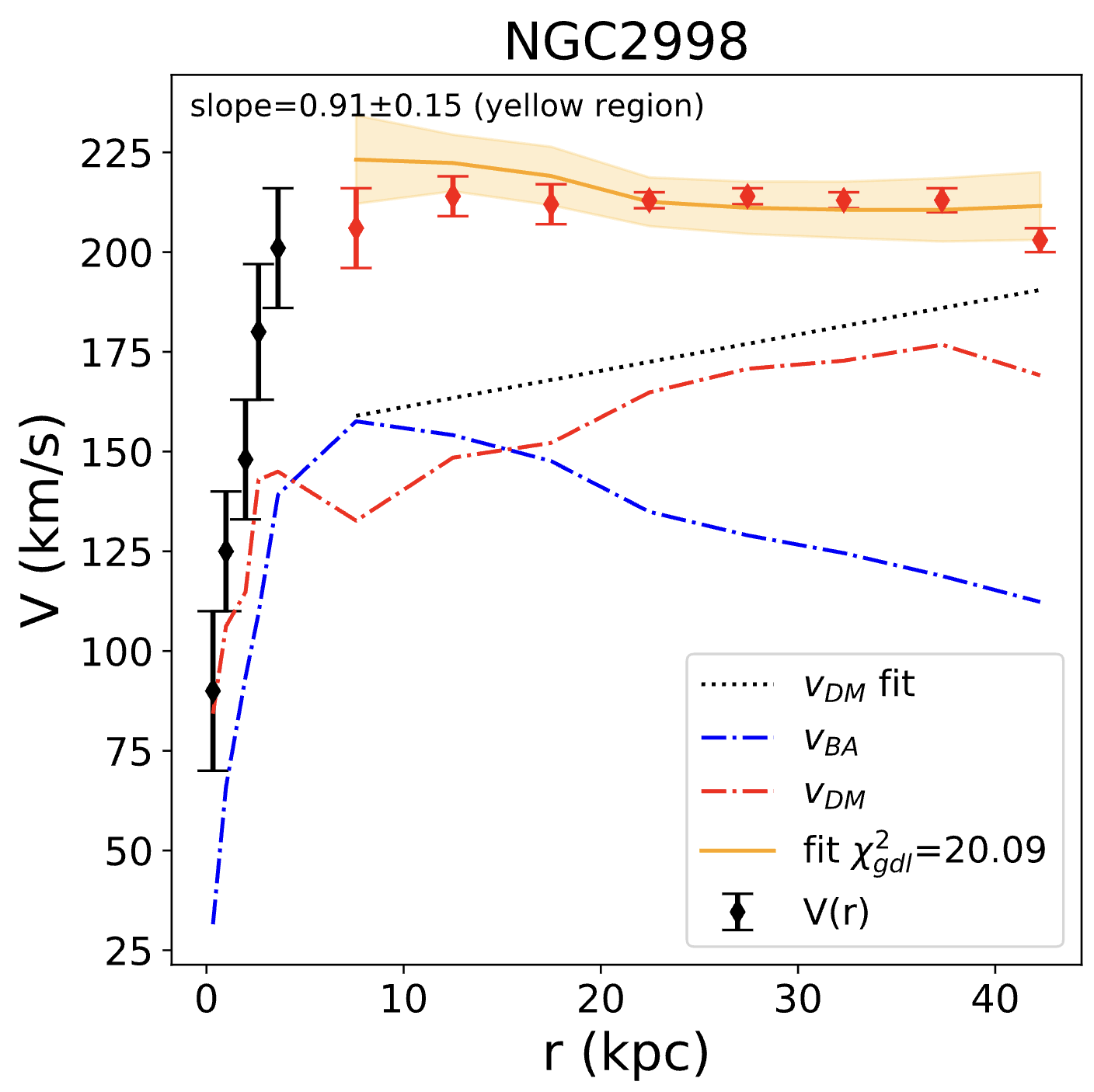}
    \includegraphics[width=0.45\linewidth]{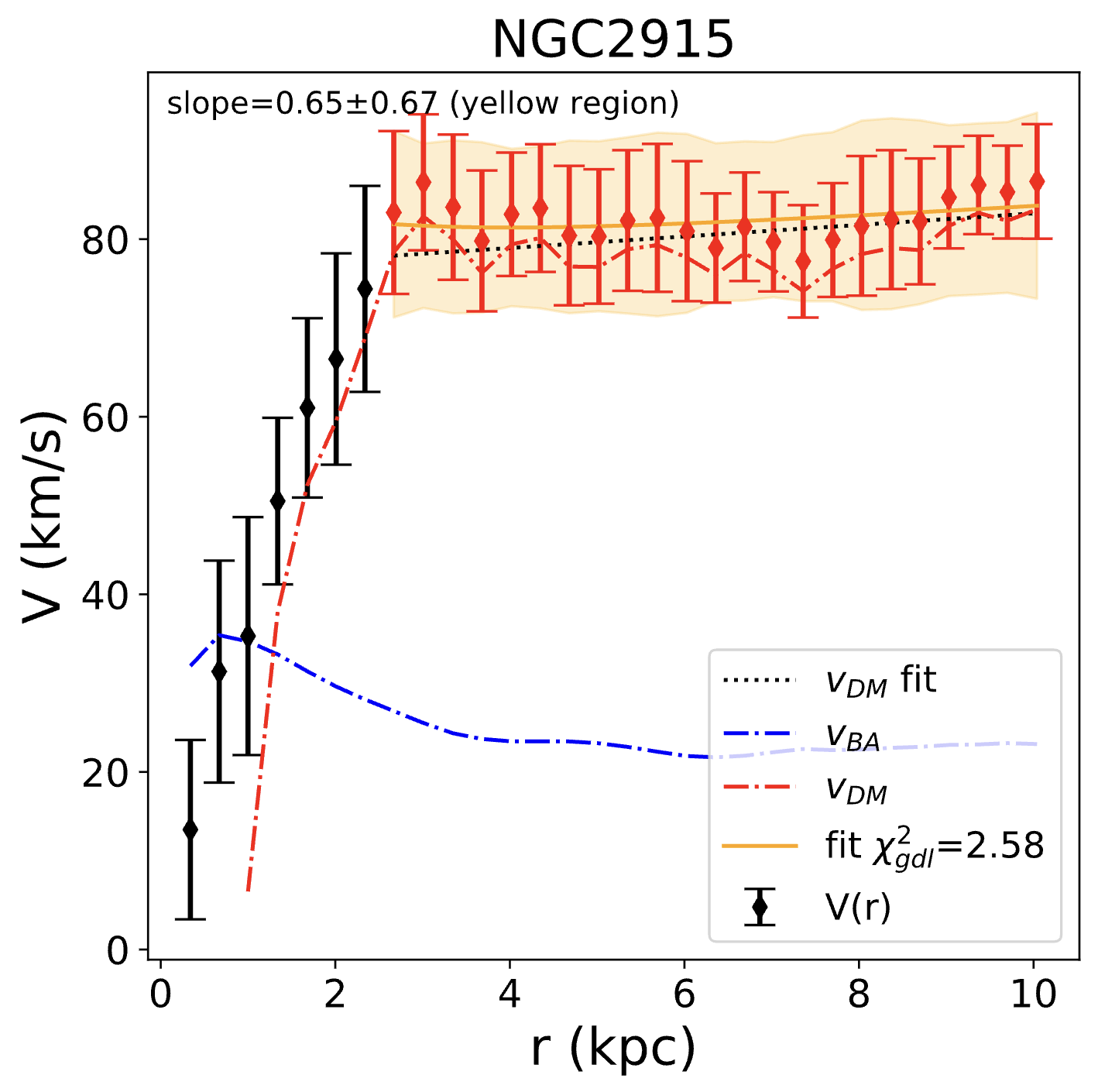}
    \includegraphics[width=0.45\linewidth]{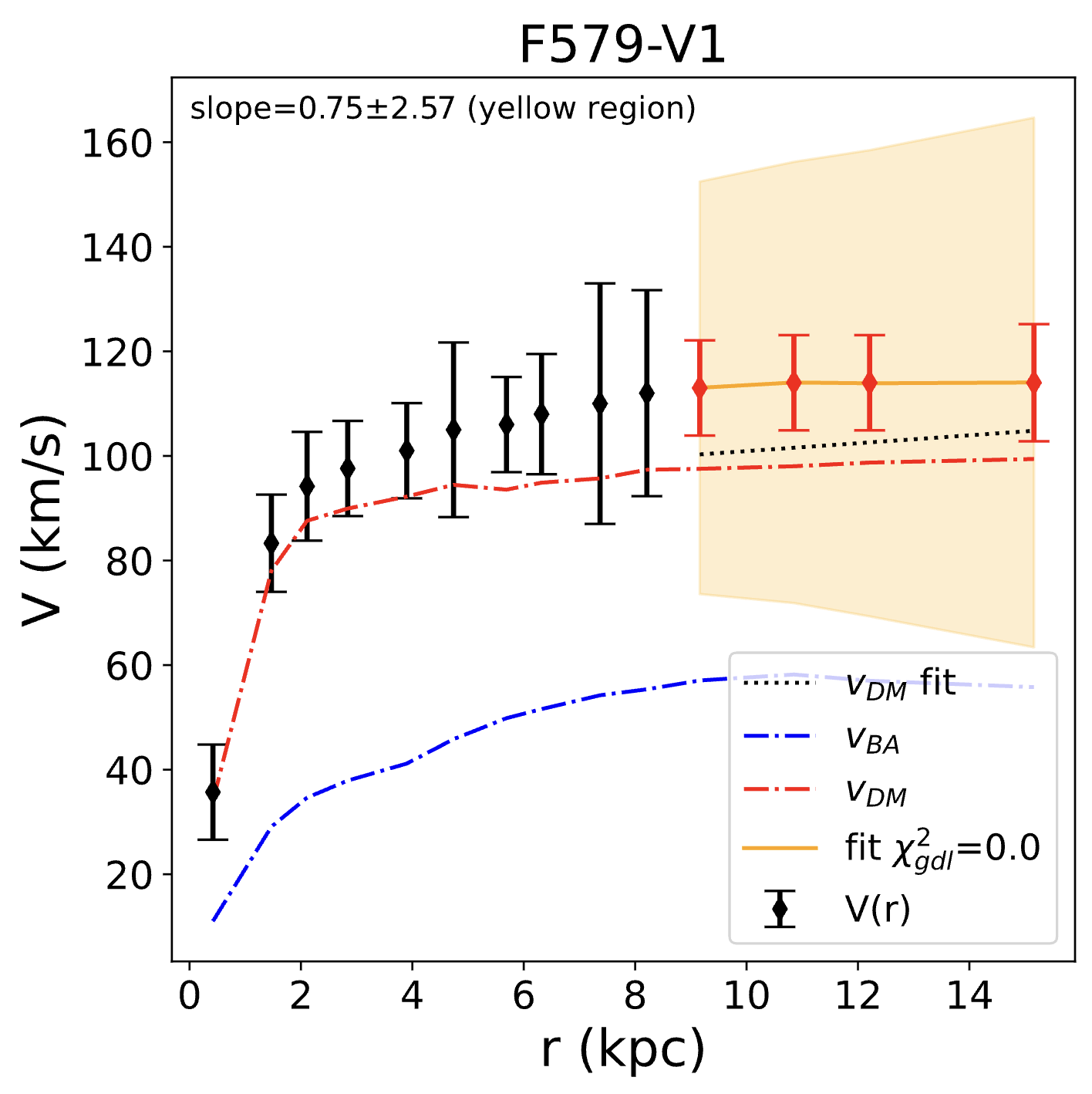}
    \includegraphics[width=0.45\linewidth]{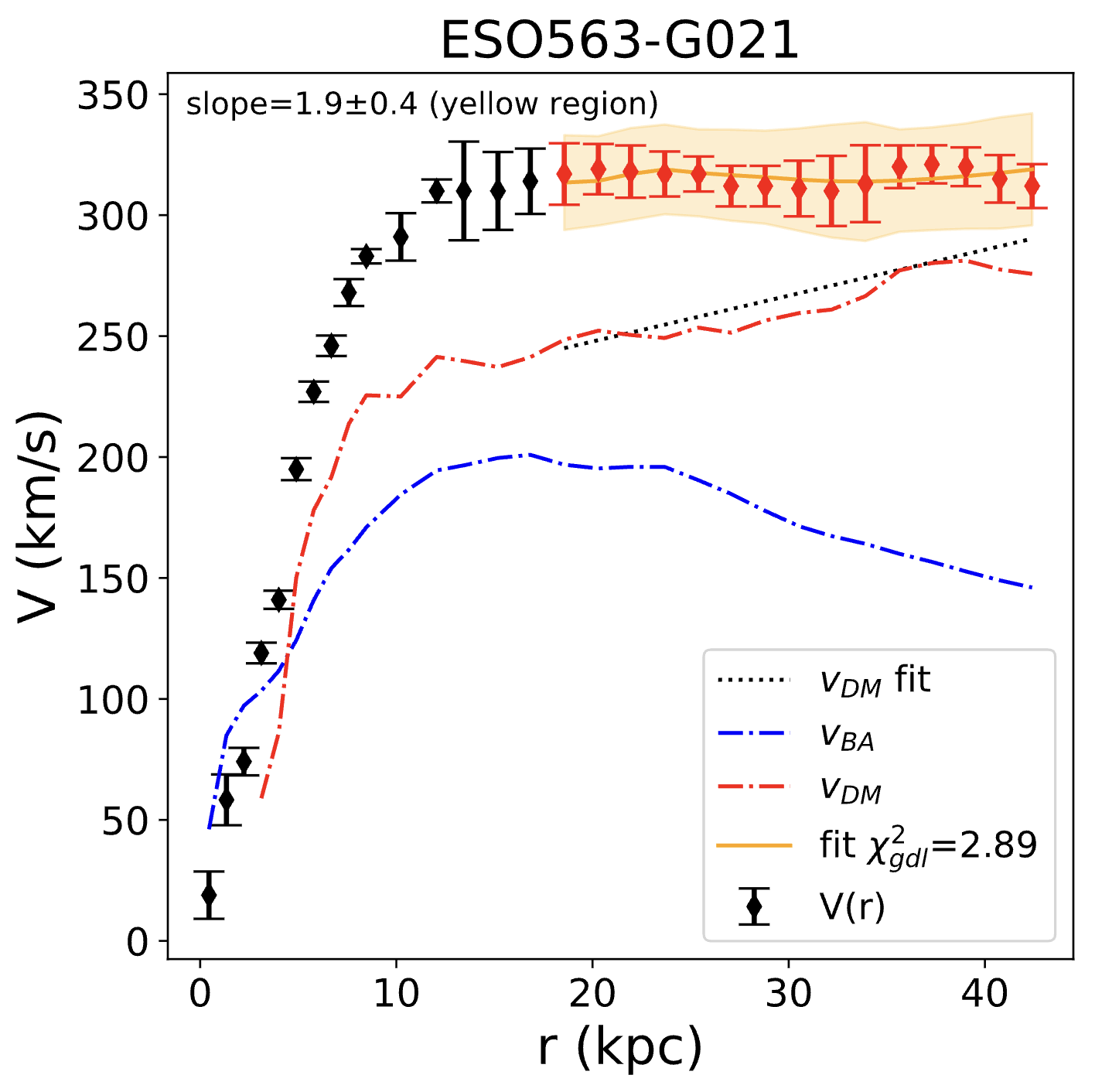}
    \includegraphics[width=0.45\linewidth]{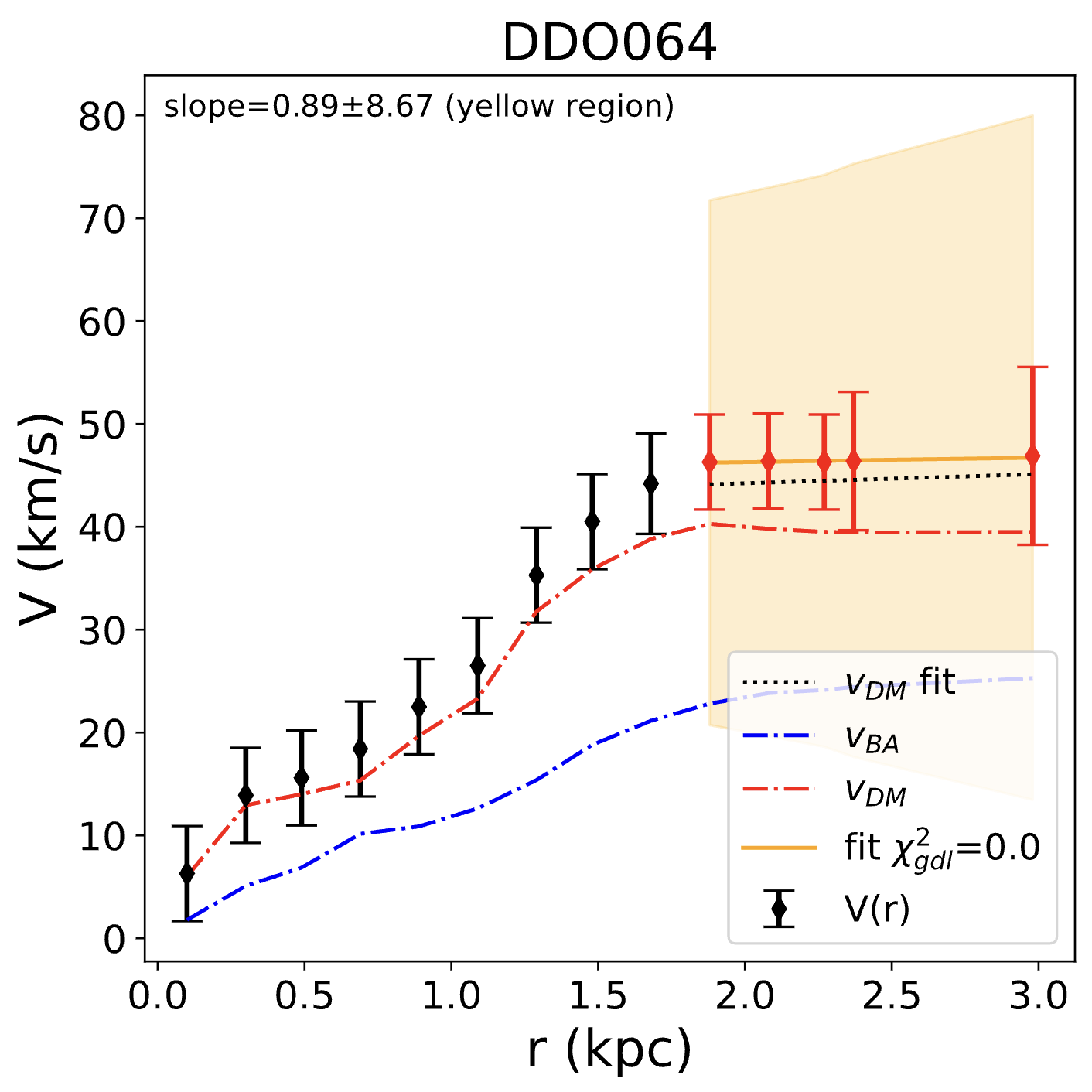}
    \caption{SPARC rotation curves with slightly positive $V'(r)_{DM}$ slope at large $r$.}
    \label{fig:examples1}
\end{figure}

\begin{figure}[h!]
    \centering
    \includegraphics[width=0.45\linewidth]{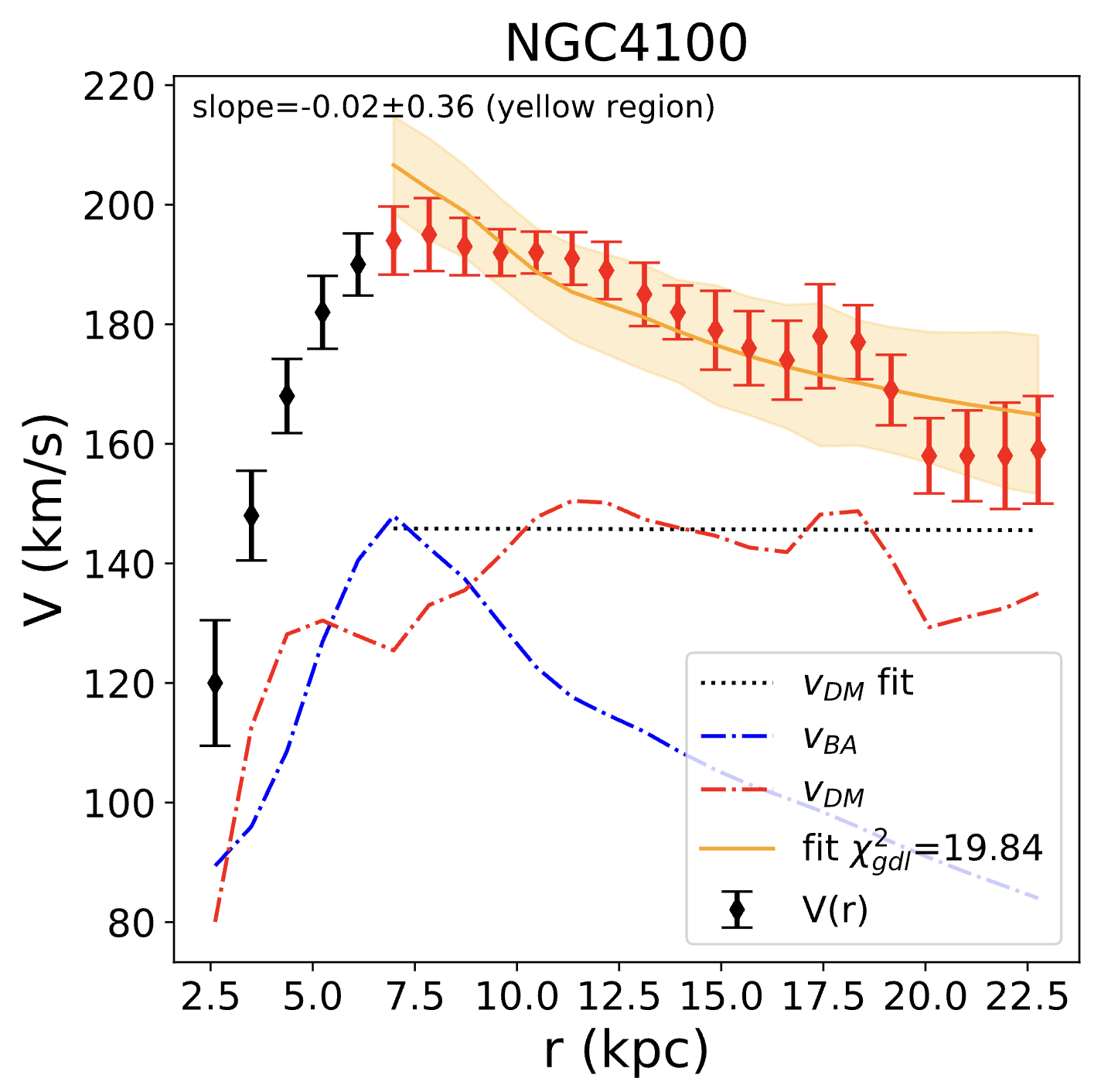}
    \includegraphics[width=0.45\linewidth]{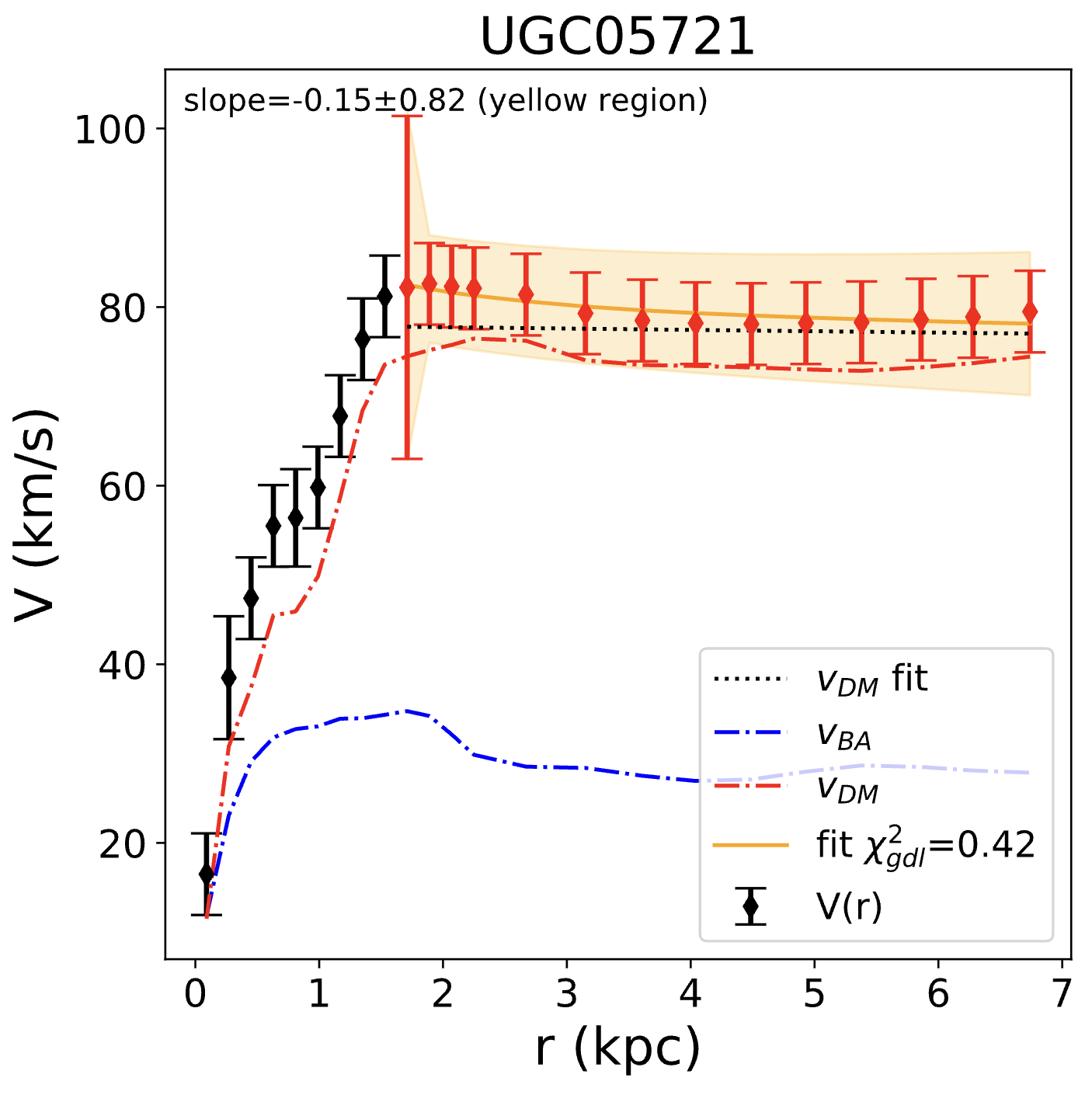}
    \includegraphics[width=0.45\linewidth]{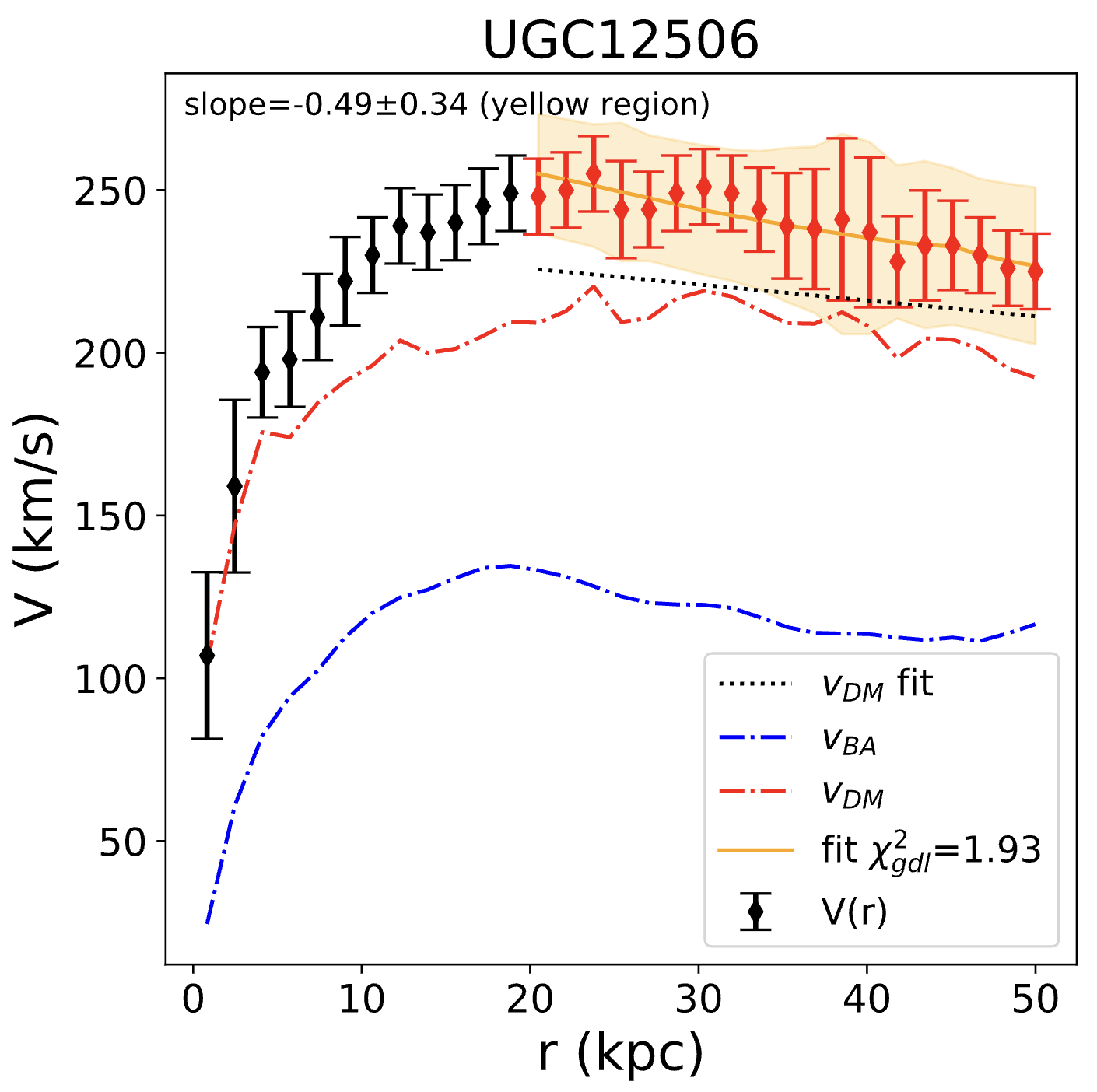}
    \includegraphics[width=0.45\linewidth]{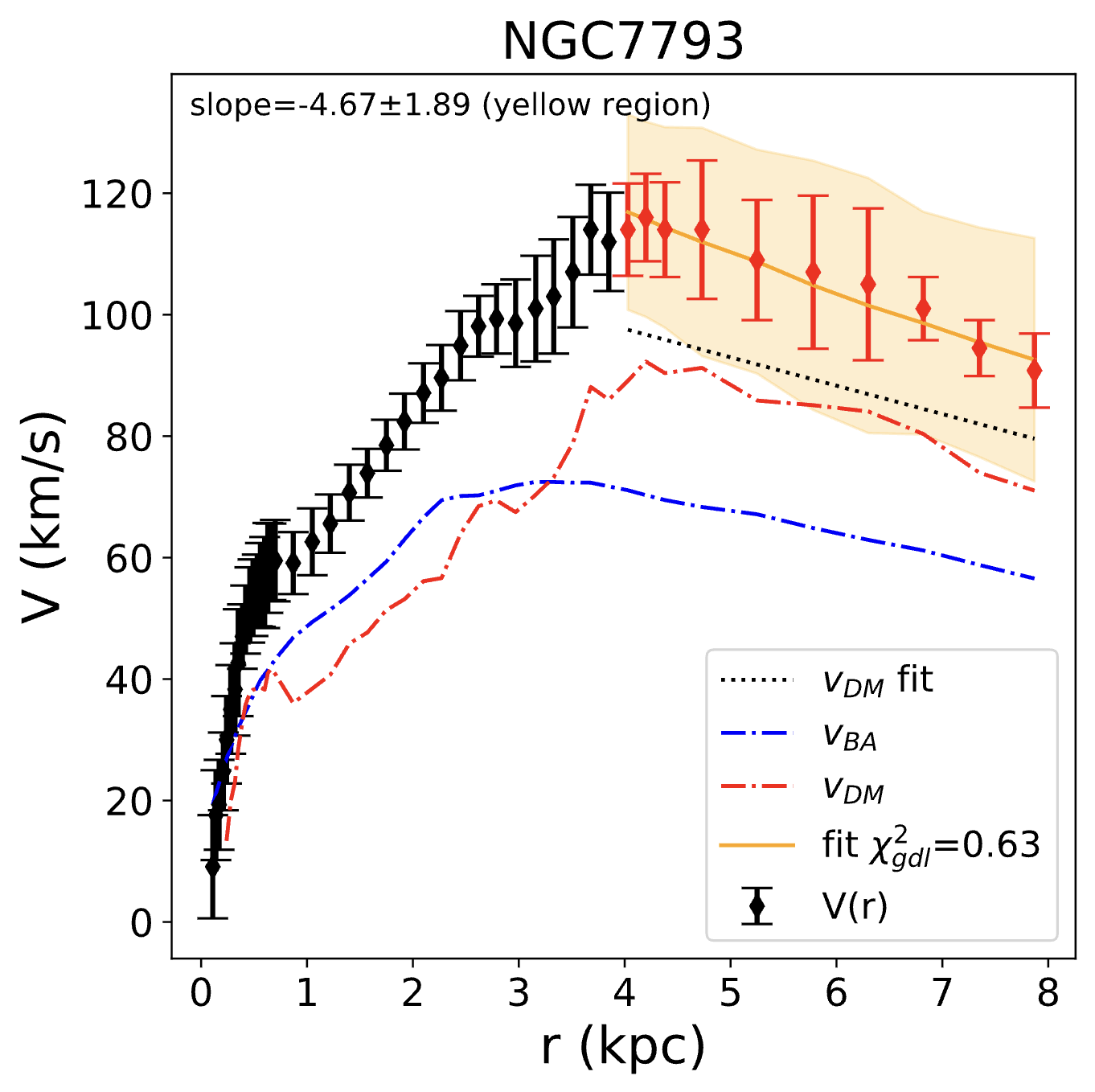}
    \includegraphics[width=0.45\linewidth]{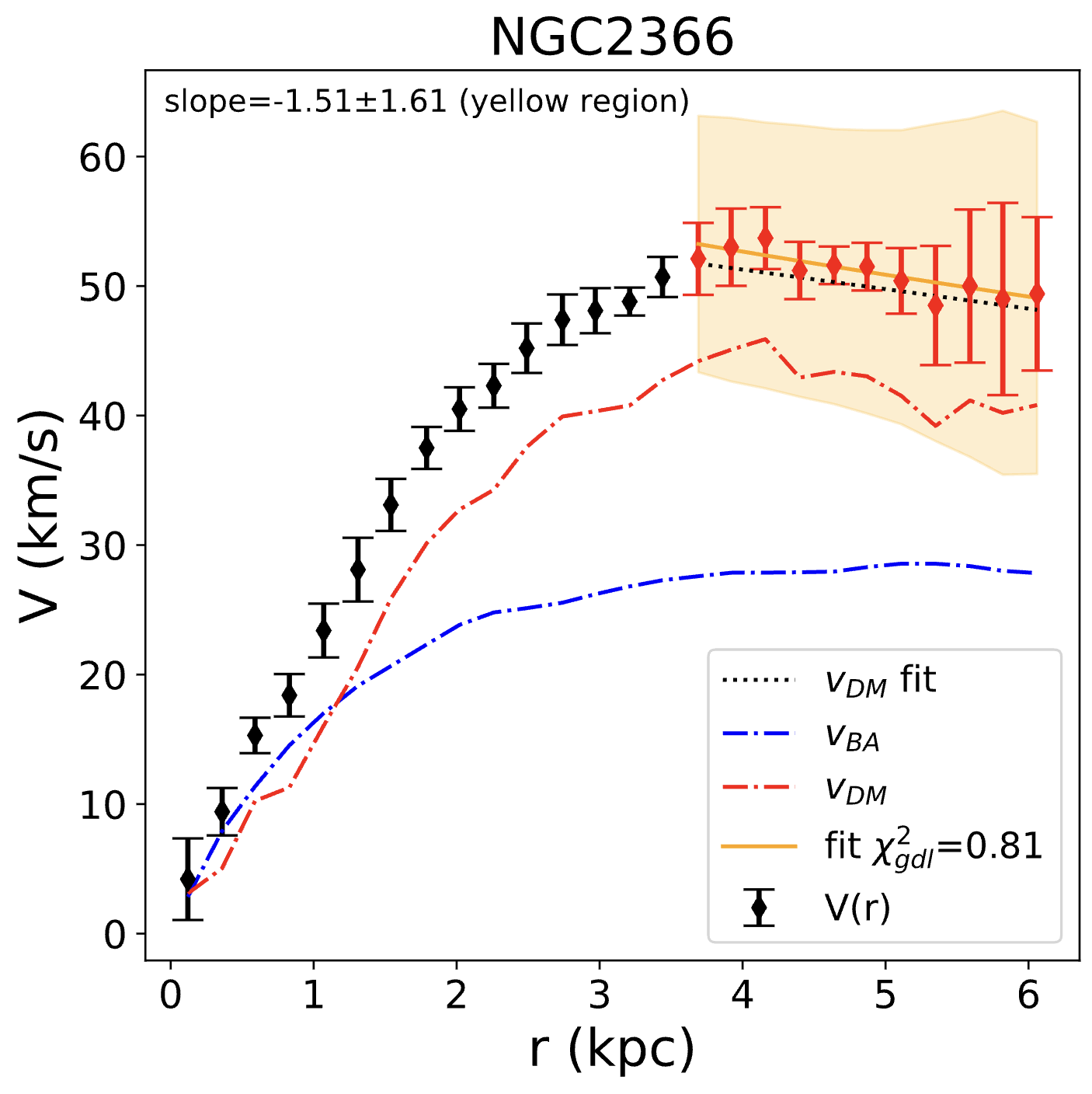}
    \includegraphics[width=0.45\linewidth]{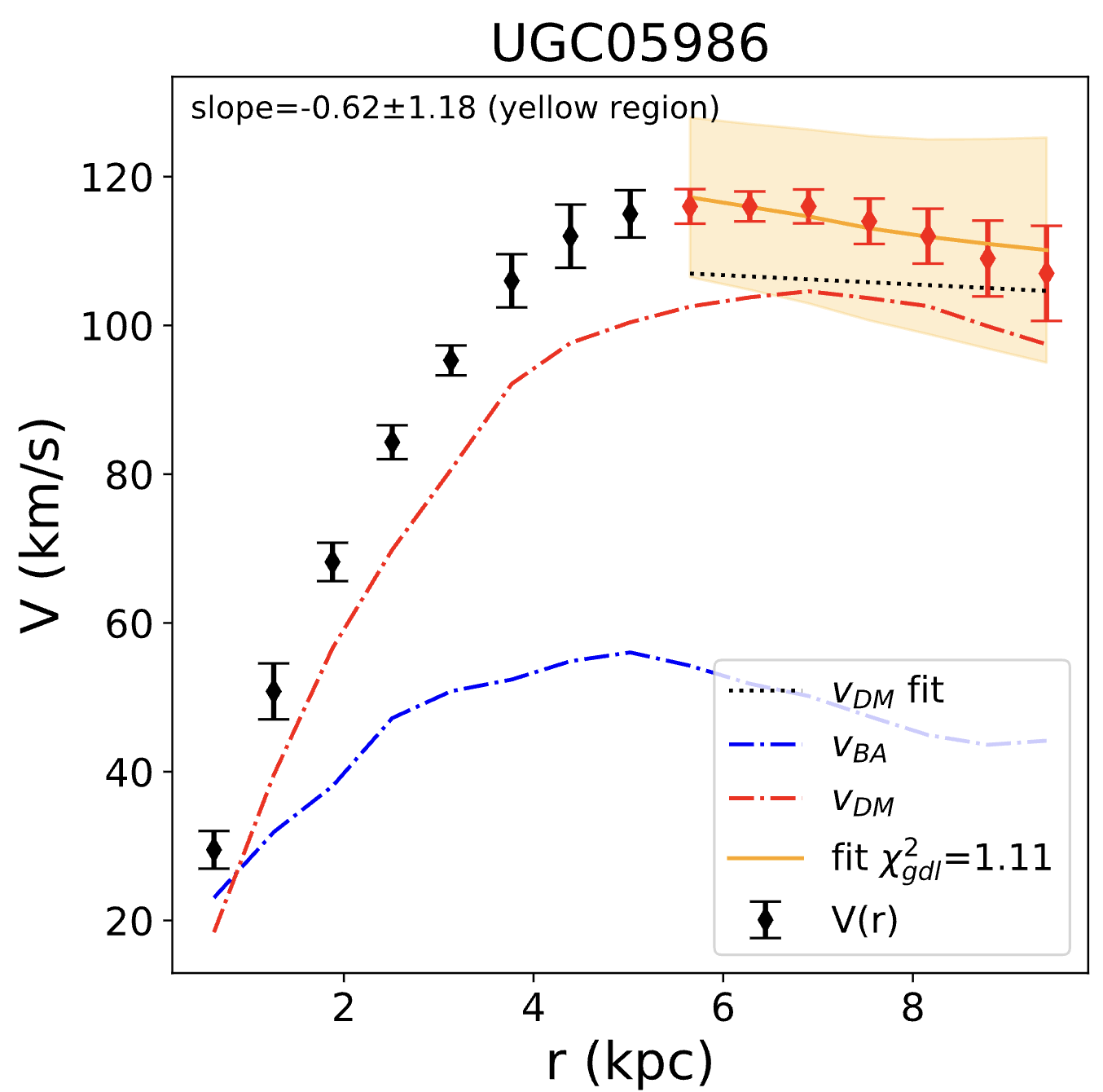}

    \caption{SPARC rotation curves with slightly negative $V'(r)_{DM}$ slope at large $r$.}
    \label{fig:examples2}
\end{figure}

%%%%%%%%%%%%%%%%%%%%%%%%%%%%%%%%%%%%%%%%%%%%%%%%%%
In Figures~\ref{fig:examples1} and~\ref{fig:examples2} we show some examples of the fitted SPARC rotation curves.  

Focusing on NG3621, the reader will see that the baryons follow Keplerian decline all the way from 4 to 17 kpc, so that it is DM that causes the slope to be flat or slightly rising; the same happens to NGC2998 with just a change of scale (between 9 and 40 kpc) and ESO653-G021 (between 15 and 45 kpc); the baryon component is negligible for NGC2915; small for F579-V1; and the only case in figure A5 which makes us entertain serious doubts is DDO064. We believe that the worry that baryons are having an outsize influence on the rotation curve, while not absurd, is not the leading physics at those large radii. There is indeed one built-in bias in the method of using $V(r)$ to learn about DM, because the tracers are stars or Hydrogen gas, and therefore the measurements are barely possible once the baryon component is completely negligible: there need to be some baryons to make these measurements.

The best-fit linear behaviour to the (DM contribution to) the $V(r)$ large-$r$ tail, the main object of study of this article, is shown as a dotted black line.
The best fit total rotation curve is the solid line (yellow online) with shaded error regions.
The points which pass the tail criterion of Eq.~(\ref{StartCondition}) and are thus the fittable data are those immersed in the shaded area (and are displayed red online), with the ones at lower $r$ not considered part of the tail.  

Finally, two more lines, dashed-dotted, are added. The usually lowest one (blue online) shows the baryon contribution to the rotation curve calculated with Eq.~(\ref{eq.2}), and the upper one (red online) the derived DM rotation curve, from Eq.~(\ref{eq.1}). Both are given for the entire $r$ range and not only for the tail, for ease of comparison with the overall data.

\begin{figure}[h!]
    \centering
    \includegraphics[width=0.45\linewidth]{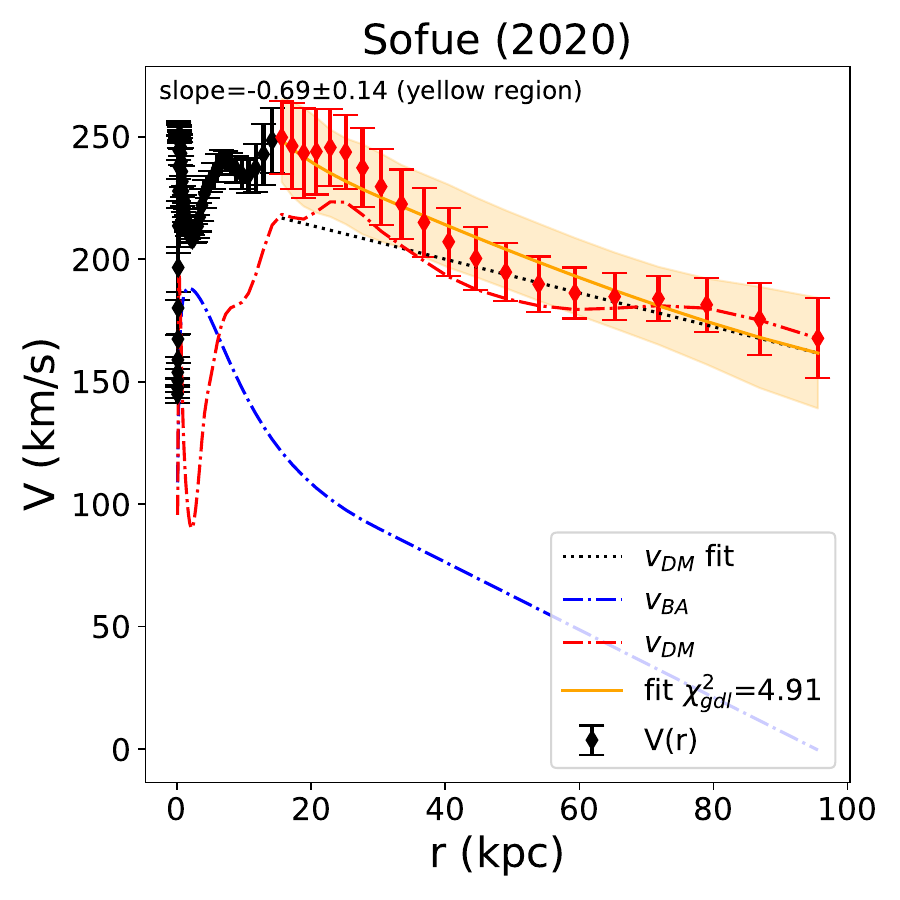}
    \includegraphics[width=0.45\linewidth]{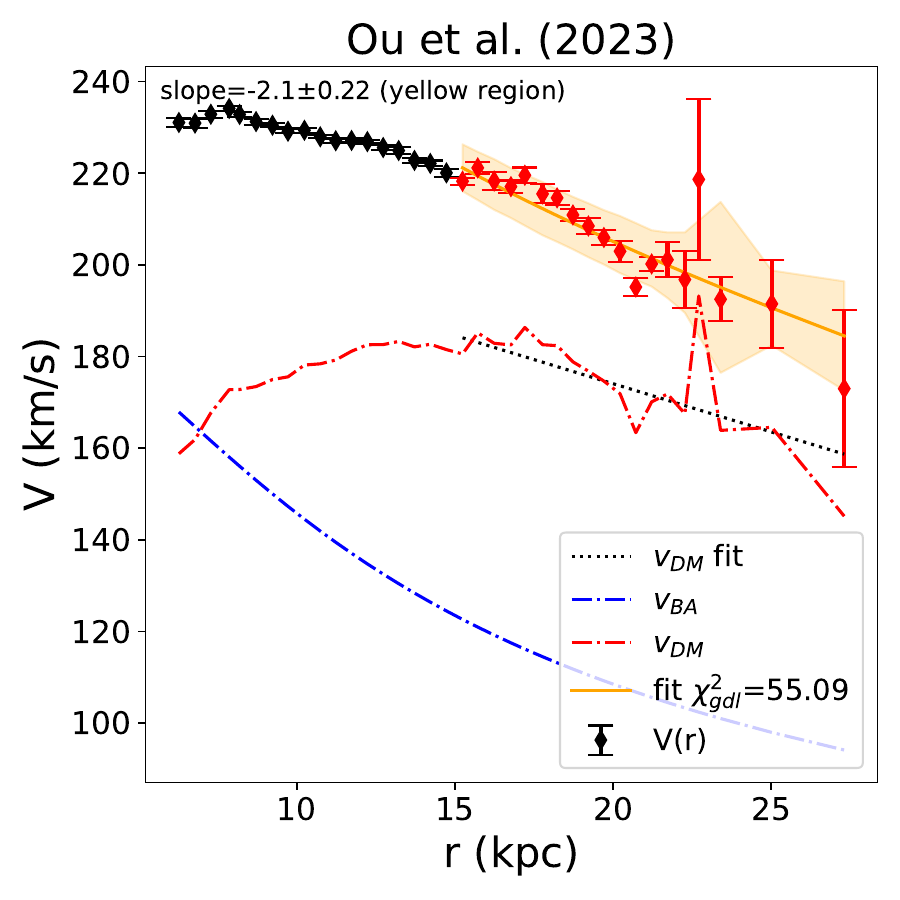}
    \includegraphics[width=0.45\linewidth]{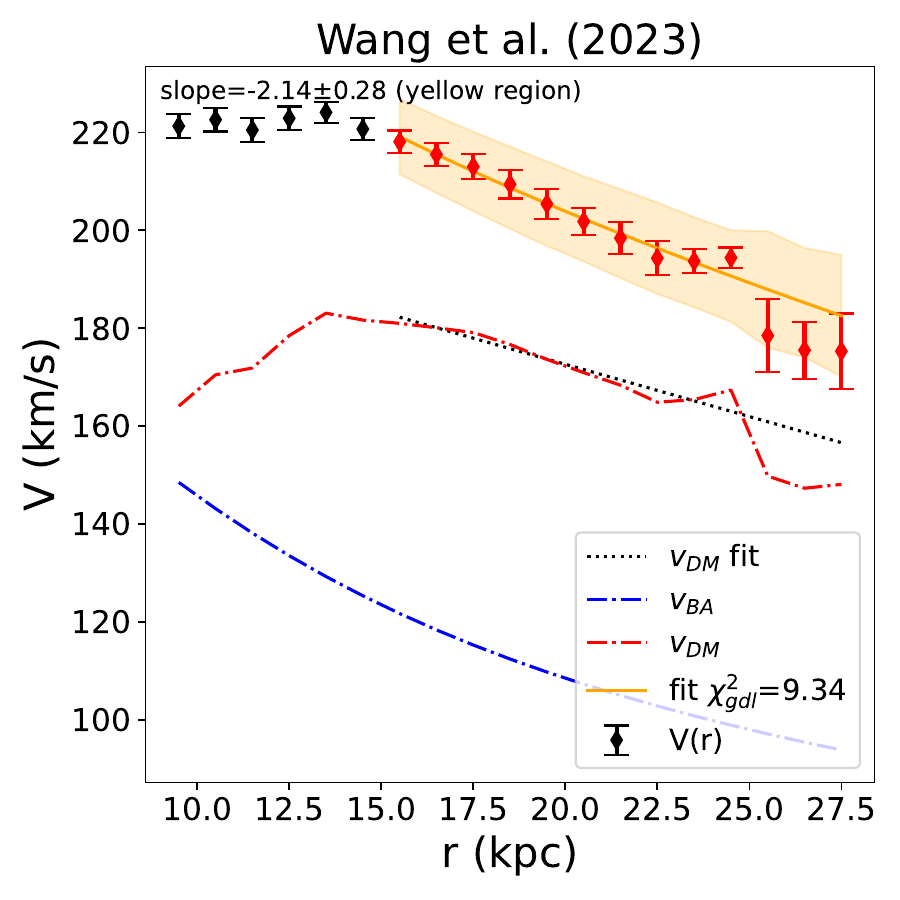}
    \includegraphics[width=0.45\linewidth]{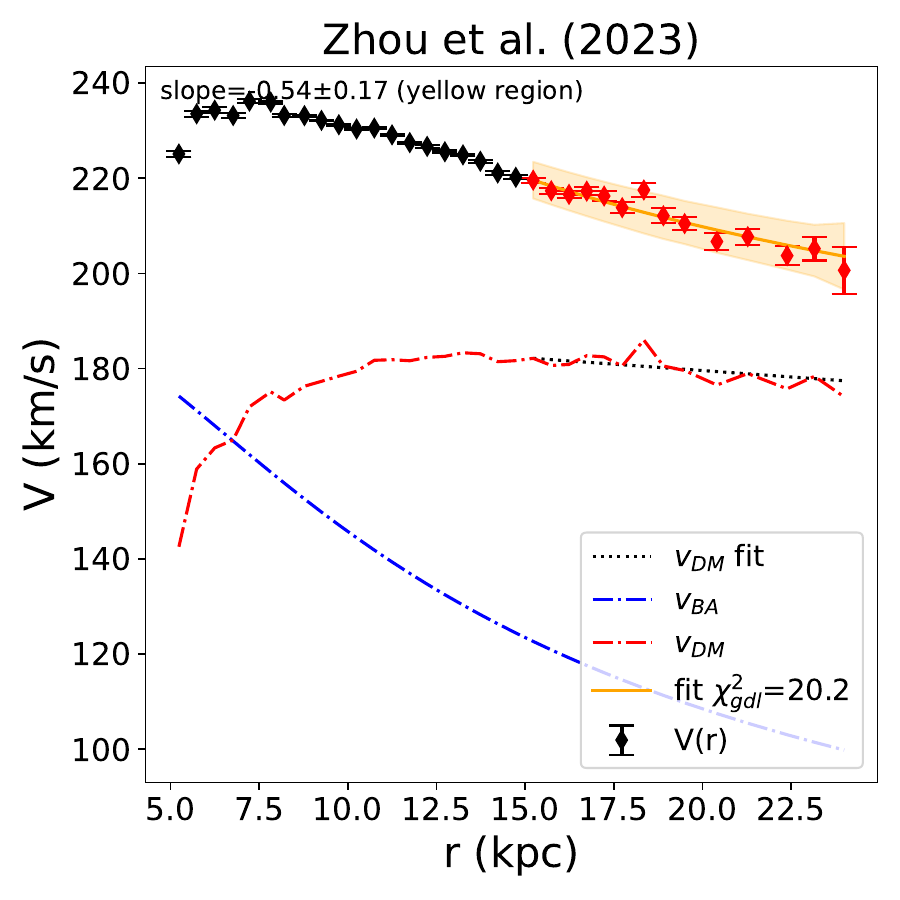}
    \includegraphics[width=0.45\linewidth]{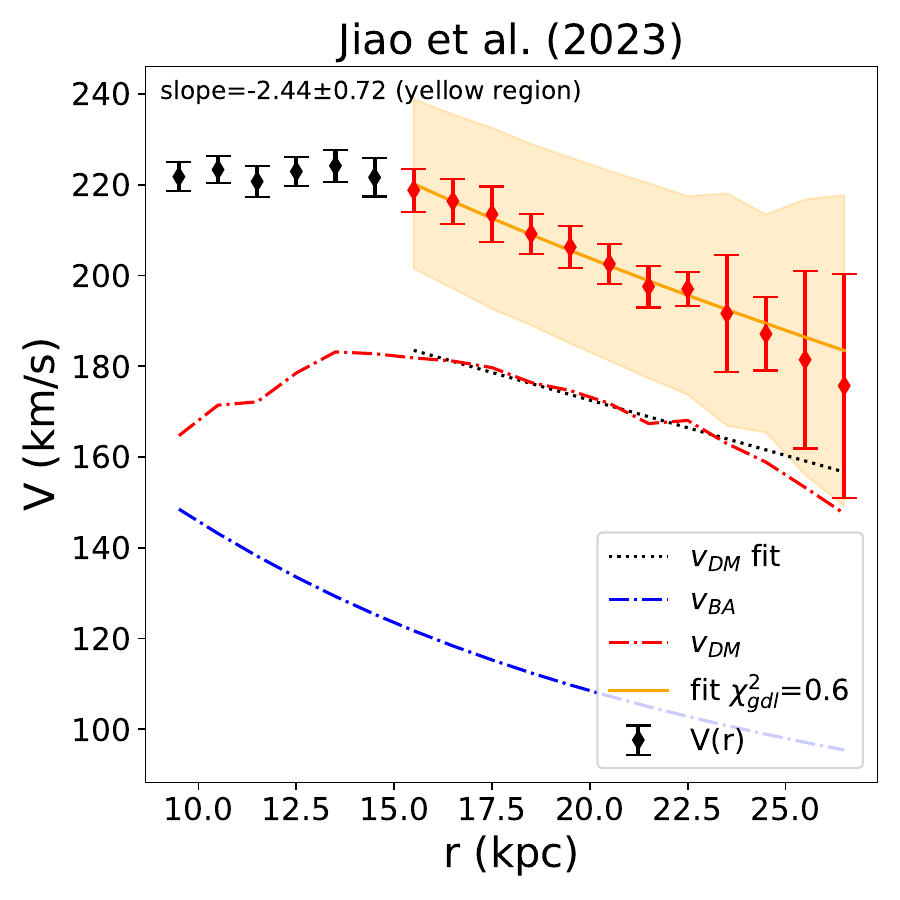}
    \includegraphics[width=0.45\linewidth]{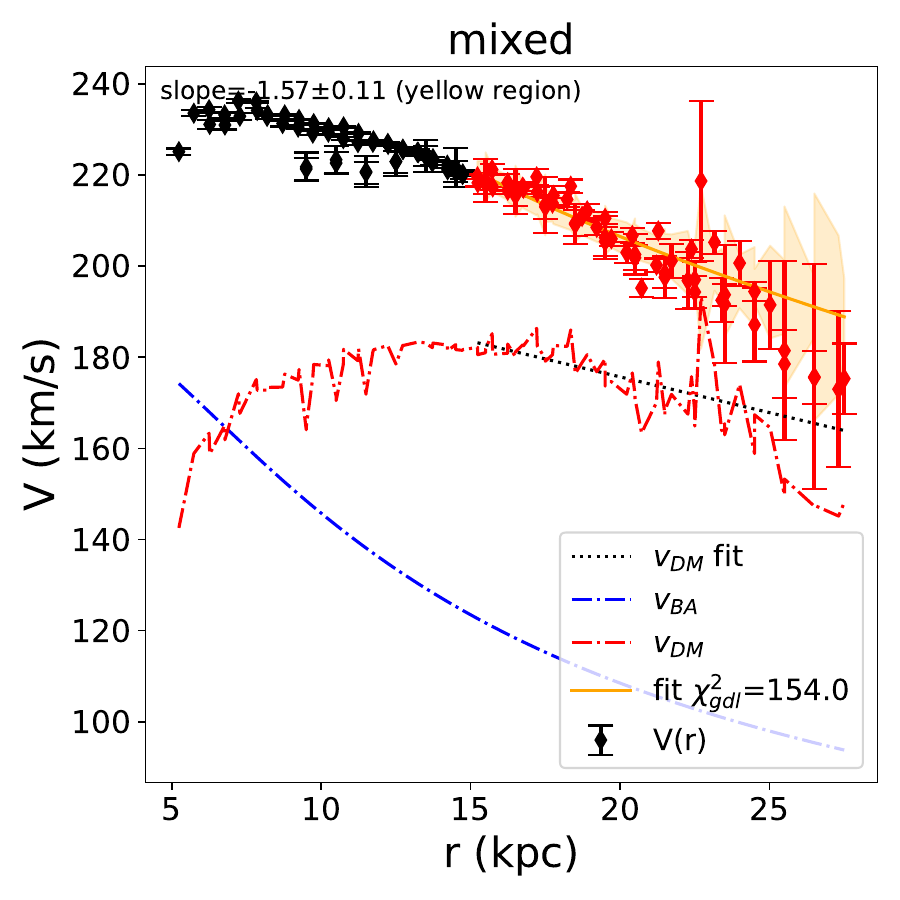}

    \caption{ Different MW rotation curve measurements and their respective fits with the DM-baryon separation based on the baryon model of~\citep{ou}. Our galaxy does show a clear decline in $V(r)$, with the newer data sets suggesting it to start at 15 kpc and the older compilation leaning towards 30 kpc.}
    \label{fig:appendix_MW}
\end{figure}

As can be seen, most are compatible with zero terminal slope (given the uncertainty on $A=V'(r)$). Also, we see that generally speaking, the sensitivity to the choice of Eq.~(\ref{StartCondition}) is not a major worry but rather a correction.
Figure~\ref{fig:appendix_MW} then displays the modern MW data.

The rest of the plots to complete the set of all fit galaxies can be found in the supplementary material.

\end{document}